\documentclass[11pt]{article}
\usepackage[left=1in,top=1in,right=1in,bottom=1in,head=.1in,nofoot]{geometry}

\usepackage{setspace,url,bm,amsmath} 

\usepackage{titlesec} 
\titlelabel{\thetitle.\quad} 

\usepackage{graphicx} 
\usepackage{bbm}
\usepackage{latexsym}
\usepackage{caption}
\usepackage[margin=20pt]{subcaption}
\usepackage{hyperref}
\usepackage{enumerate}
\usepackage{lscape}
\usepackage{listings}
\usepackage{multirow}
\usepackage[table]{xcolor}

\newcommand{\GG}[1]{}

\usepackage{amsthm}
\usepackage{comment}
\theoremstyle{definition}
\newtheorem{assumption}{Assumption}
\newtheorem*{theorem*}{Theorem}
\newtheorem{theorem}{Theorem}
\newtheorem{proposition}{Proposition}

\newtheorem{example}{Example}

\newtheorem{definition}{Definition}

\newtheorem*{corollary*}{Corollary}

\usepackage{natbib} 
\bibpunct{(}{)}{;}{a}{}{,} 

\usepackage{etoolbox} 
\apptocmd{\sloppy}{\hbadness 10000\relax}{}{} 

\usepackage{color}
\usepackage{listings}

\def\ind{\begin{picture}(9,8)
         \put(0,0){\line(1,0){9}}
         \put(3,0){\line(0,1){8}}
         \put(6,0){\line(0,1){8}}
         \end{picture}
        }

\def\var{\text{var}}
\def\cov{\text{cov}}
\def\sumn{\sum_{i=1}^n}

\def\pr{\textup{pr}}
\def\T{{ \mathrm{\scriptscriptstyle T} }}

\begin{document}
\onehalfspacing
\title{\bf \Large Flexible sensitivity analysis for causal inference in observational studies subject to unmeasured confounding} 
\author{Sizhu Lu and Peng Ding \footnote{ Department of Statistics, University of California, Berkeley, CA 94720, USA (Emails: \url{sizhu_lu@berkeley.edu} and \url{pengdingpku@berkeley.edu})}
}
\date{}
\maketitle

\begin{abstract}
Causal inference with observational studies often suffers from unmeasured confounding, yielding biased estimators based on the unconfoundedness assumption. Sensitivity analysis assesses how the causal conclusions change with respect to different degrees of unmeasured confounding. Most existing sensitivity analysis methods work well for specific types of statistical estimation or testing strategies. We propose a flexible sensitivity analysis framework that can deal with commonly-used inverse probability weighting, outcome regression, and doubly robust estimators simultaneously. It is based on the well-known parametrization of the selection bias as comparisons of the observed and counterfactual outcomes conditional on observed covariates. It is attractive for practical use because it only requires simple modifications of the standard estimators. Moreover, it naturally extends to many other causal inference settings, including the causal risk ratio or odds ratio, the average causal effect on the treated units, and studies with survival outcomes.  We also develop an R package \texttt{saci} to implement our sensitivity analysis estimators.

\end{abstract}

\section{Introduction}
\label{sec::introduction}

Learning causal effects is of great importance in empirical research, which, however,  presents a persistent challenge for researchers. When randomized control trials are not feasible, we need to learn from the observational data to conduct causal inference. The standard method for identifying causal parameters, such as the average causal effect, hinges upon the unconfoundedness assumption. Unconfoundedness postulates that the potential outcomes are conditionally independent of the treatment assignment given the observed covariates. However, unconfoundedness is untestable and necessitates the measurement of all confounding factors associated with the treatment and outcome. The presence of unmeasured confounders yields bias in estimating the true causal effect, even with a substantial sample size. 

Motivated by the concern of unmeasured confounding, sensitivity analysis becomes an attractive alternative to assess the impact of the unobserved confounders on the causal estimates. \citet{Rosenbaum::1983JRSSB}, \citet{lin1998assessing} and \citet{imbens2003sensitivity} built parametric models to assess the impact of the unobserved confounder on the estimation of the average causal effect. \citet{rosenbaum1987sensitivity} proposed a sensitivity analysis framework to test the sharp null hypothesis of no unit-level causal effects in matched observational studies, with a recent application by \cite{zubizarreta2013effect}. \citet{cornfield1959smoking} and \citet{ding2016sensitivity} derived inequalities to quantify the strength of the unmeasured confounding in order to explain away observed causal estimates based on risk ratios, which motivated the concept of E-value for observational research for causal inference \citep{vanderweele2017sensitivity}. \citet{zhao2019sensitivity} and \citet{dorn2022sharp} developed sensitivity analysis methods for the inverse propensity score weighting estimator. 

The methods mentioned above for sensitivity analysis are useful for specific statistical estimation or testing strategies. However, they may be limited in their ability to simultaneously deal with various types of estimators, such as outcome regression-based estimators, inverse propensity score weighting estimators, and doubly robust estimators. To develop a comprehensive approach, we propose a unified framework for sensitivity analysis adaptable to different estimators. Our framework utilizes the sensitivity parameters defined as the ratio between the conditional means of potential outcome in two treatment groups given the observed covariates \citep{robins1999association}. We provide nonparametric identification formulas for the average causal effect given the specified sensitivity parameters. Our framework facilitates sensitivity analysis for various standard estimators by providing closed-form modifications. Furthermore, we furnish practical guidance on calibrating and specifying sensitivity parameters to interpret the sensitivity analysis results. 

\subsection{A running example}
\label{subsec::motivating_example}
Consider the causal impact of smoking on individuals' health as a motivating example. Randomly assigning individuals to smoke is neither practical nor ethical, thus we need to rely on observational data to learn the causal effect of interest. Specifically, the observational study in \cite{bazzano2003relationship} explored whether cigarette smoking positively affects homocysteine levels, a known risk factor for cardiovascular disease. Drawing upon observations from the U.S. National Health and Nutrition Examination Survey (NHANES) 2005--2006, the study compares homocysteine levels between smokers and non-smokers while adjusting for observed covariates like age, gender, race, and body mass index (BMI) under the unconfoundedness assumption. 
However, in the presence of suspected underlying factors associated with both increased probability of smoking and elevated homocysteine levels, such as certain gene expressions, estimates adjusted solely for observed covariates are prone to bias. In such instances, conducting sensitivity analyses becomes essential to discern how the estimated causal effect changes in response to the presence of unmeasured confounders. Such analyses help determine the degree to which we need to deviate from the unconfoundedness assumption to attribute the estimated positive effect to unobserved confounders. We will revisit this example in Section~\ref{subsec::application} to illustrate our proposed methods.

\subsection{Organiztion of the paper}
The remainder of the paper is organized as follows. In Section~\ref{sec::unconfounded}, we review causal inference in observational studies under the unconfoundedness assumption. In Section~\ref{sec::sensitivity_analysis}, we introduce our sensitivity analysis framework and provide identification and estimation procedures. In Section~\ref{sec::calibration}, we provide practical suggestions for calibration and specification of the sensitivity parameters. In Section~\ref{sec::application_simulation}, we revisit the motivating example, apply the proposed method to this real-world application, and conduct a simulation study to evaluate the finite sample performance of our proposed methods. In Section~\ref{sec::extensions}, we discuss alternative sensitivity parameters as well as the extensions to other causal inference settings. In Section~\ref{sec::discussion}, we conclude. In the supplementary material, we present additional technical details.

\section{Review of causal inference in unconfounded observational studies} 
\label{sec::unconfounded}

Causal inference in observational studies is a central task in many disciplines. The potential outcomes framework is a leading approach to causal inference in statistics. Let $Y_i(1)$ and $Y_i(0)$ denote the potential outcomes if unit  $i$ receives the treatment and control, respectively. Let $Z_i$ denote the binary treatment, with $Z_i=1$ if unit $i$ receives the treatment and $Z_i=0$ if unit $i$ receives the control, respectively. Let
$Y_i = Y_i(Z_i) =  Z_i Y_i(1) + (1-Z_i) Y_i(0)$ denote the observed outcome, and $X_i$ denote the observed covariates for unit $i$. 
We focus on the standard setting with independent and identically distributed $\{ X_i, Z_i, Y_i(1), Y_i(0) : i =1, \ldots, n \}$, and will drop the index $i$ henceforth for the description of population quantities. We start with the average causal effect
$
\tau = E\{ Y(1) - Y(0) \} .
$
By the law of total probability, it decomposes into 
\begin{eqnarray} 
\tau &=& \left[  E(Y\mid Z=1) \pr(Z=1) + E\{ Y(1)\mid Z=0 \} \pr(Z=0) \right]  \nonumber  \\
&&- \left[   E\{ Y(0)\mid Z=1\} \pr(Z=1) + E(Y\mid Z=0)  \pr(Z=0)  \right]. \label{eqn::decompositon-ace}
\end{eqnarray} 
In \eqref{eqn::decompositon-ace}, the fundamental difficulty is to estimate the counterfactual means $E\{ Y(1)\mid Z=0 \} $ and $E\{ Y(0)\mid Z=1\}$, which are not identifiable without further assumptions. 
\citet{rubin1978bayesian} and \citet{rosenbaum1983central} proposed the following unconfoundedness assumption as a sufficient condition to identify $\tau$. 

\begin{assumption}
[unconfoundedness] 
\label{assume::unconfoundedness}
$Z \ind \{ Y(1), Y(0)  \} \mid X$.
\end{assumption}

Under Assumption \ref{assume::unconfoundedness}, we can prove two identification formulas for $\tau$ which write our causal parameter of interest in terms of the observed data distribution: 
\begin{eqnarray} 
\tau &=&   E\{ \mu_1(X)  - \mu_0(X) \} \ =\  E\left\{   \frac{ZY}{e(X)} -  \frac{ (1-Z)Y}{1-e(X)}    \right\}, \label{eqn::identification-tau}
\end{eqnarray}
where $\mu_1(X) = E(Y\mid Z=1, X)$ and $\mu_0(X) = E(Y\mid Z=0, X)$ are the conditional means of the outcome given covariates under treatment and control, respectively, and $e(X)  = \pr(Z= 1\mid X)$ is the conditional mean of the treatment given covariates, also called the propensity score \citep{rosenbaum1983central}. The formulas in~\eqref{eqn::identification-tau} rely on the overlap assumption: $0<e(X)<1$. We implicitly assume it, since the focus of this paper is on the unconfoundedness assumption.

The two plug-in estimators corresponding to the two identification formulas for $\tau$ in~\eqref{eqn::identification-tau} are the outcome regression estimator and the Horvitz--Thompson-type inverse propensity score weighting estimator, respectively:
\begin{eqnarray*} 
\hat\tau^\textup{reg} &=& n^{-1} \sum_{i=1}^n  \{ \hat \mu_1(X_i)  - \hat \mu_0(X_i)\} ,\quad \hat\tau^\textup{ht} \ =\  n^{-1} \sum_{i=1}^n \left\{   \frac{Z_i Y_i }{ \hat e(X_i)} -  \frac{ (1-Z_i)Y_i}{1- \hat e(X_i)}    \right\},
\end{eqnarray*}
with $\hat e(X_i)$ and $\hat \mu_z(X_i)$ signifying fitted propensity score and outcome models. Moreover, we can construct the doubly robust estimator by combining both models \citep{bang2005doubly}:
\begin{equation} 
\hat\tau^\textup{dr} \ =\  \hat\tau^\textup{reg} 
+n^{-1} \sum_{i=1}^n \left\{   \frac{Z_i \check Y_i }{ \hat e(X_i)} -  \frac{ (1-Z_i) \check Y_i }{1- \hat e(X_i)}    \right\} \ =\ \hat\tau^\textup{ht}  
- n^{-1} \sum_{i=1}^n \check Z_i  \left\{   \frac{ \hat \mu_1(X_i) }{ \hat e(X_i)}    + \frac{ \hat \mu_0(X_i) }{ 1-\hat e(X_i)}   \right\} , \label{eqn::doubly-robust-two-forms}
\end{equation} 
where $\check Y_i  = Y_i - \hat\mu_{Z_i}(X_i)$ and $ \check Z_i =  Z_i-  \hat e(X_i)$ are the residuals from the outcome model and propensity score model, respectively.  The two equivalent forms in \eqref{eqn::doubly-robust-two-forms} highlight that we can view $\hat\tau^\textup{dr} $ as a modification of either $\hat\tau^\textup{reg} $ or $ \hat\tau^\textup{ht}  $: it modifies $\hat\tau^\textup{reg} $ by inverse propensity score weighted residuals, and augments $\hat\tau^\textup{ht} $ by the imputed outcomes.

\section{Sensitivity analysis with unmeasured confounding}
\label{sec::sensitivity_analysis}

\subsection{Motivation for sensitivity analysis and parametrization of unmeasured confounding}

Assumption \ref{assume::unconfoundedness} is strong and untestable. Observational studies are often plagued with unmeasured confounding, that is, the existence of some hidden variables $U$ that affect both the treatment and outcome simultaneously. Under these circumstances, we need to conduct sensitivity analyses to assess the impact of $U$ on the causal estimates. Existing methods only serve purposes in specific statistical estimation or testing strategies, as reviewed in Section~\ref{sec::introduction}. To provide a unified framework that can deal with the standard estimators  $\hat\tau^\textup{reg} $, $ \hat\tau^\textup{ht}  $ and $\hat\tau^\textup{dr} $ reviewed in Section \ref{sec::unconfounded} simultaneously, we adopt the following parametrization of confounding. 

\begin{definition}\label{def::sensitivity-analysis}
Define 
\begin{eqnarray*}
\frac{ E\{ Y(1)\mid Z=1, X \} }{ E\{ Y(1)\mid Z=0, X \} } = \varepsilon_1(X), \quad 
\frac{ E\{ Y(0)\mid Z=1, X \} }{ E\{ Y(0)\mid Z=0, X \} } = \varepsilon_0(X) .
\end{eqnarray*}
\end{definition}

In Definition \ref{def::sensitivity-analysis}, $\varepsilon_1(X)$ and $\varepsilon_0(X)$ are two sensitivity parameters. In sensitivity analysis, we can first fix them to obtain the corresponding estimators and then vary them within a range to obtain a sequence of estimators. Definition \ref{def::sensitivity-analysis}, as well as the theory below, allows $\varepsilon_1(X)$ and $\varepsilon_0(X)$ to depend on the covariates. For simplicity of implementation, we can further assume that they are not functions of the covariates.

The parametrization in Definition \ref{def::sensitivity-analysis} compares the observable conditional means of the potential outcomes, $E\{ Y(1)\mid Z=1, X \}$ and $E\{ Y(0)\mid Z=0, X \}$, with the corresponding counterfactual conditional means of the potential outcomes, $E\{ Y(1)\mid Z=0, X \}$ and $E\{ Y(0)\mid Z=1, X \}$, respectively. It quantifies the violation of unconfoundedness in Assumption \ref{assume::unconfoundedness}. Technically, the form of unconfoundedness in Assumption \ref{assume::unconfoundedness} is stronger than we need if the parameter of interest is $\tau$. When $\varepsilon_1(X) = \varepsilon_0(X) = 1$ in Definition \ref{def::sensitivity-analysis}, we can recover all identification and estimation results for $\tau$ in Section \ref{sec::unconfounded}. We revisit the running example in Section~\ref{subsec::motivating_example}. If suspected factors associated with both increased probability of smoking and elevated homocysteine levels exist, under Definition~\ref{def::sensitivity-analysis}, $\varepsilon_1(X) > 1$ and $\varepsilon_0(X) > 1$, leading to an upward bias of estimators of $\tau$ constructed under the unconfoundedness assumption. The sensitivity parameters in Definition~\ref{def::sensitivity-analysis} are also related to the parametrization based on the treatment selection model $\pr\{Z=z\mid X, Y(z)\}$ for $z=0,1$. We provide more discussion in Section~\ref{subsec::treatment_selection} in the supplementary material.

\citet{robins1999association} and \citet{scharfstein1999adjusting} discussed the potential use of the parametrization in Definition \ref{def::sensitivity-analysis}. Similar parametrizations also appeared in other settings for causal inference \citep{tchetgen2012semiparametric, vanderweele2014interference, ding2017principal, yang2018sensitivity, jiang2020multiply}. The confounding function in \citet{kasza2017assessing} also adopts ratios between the marginal means, $E\{Y(1)\mid Z=1\}/E\{Y(1)\mid Z=0\}$, in the special case of binary outcomes, without specifying it as a flexible function of the observed covariates. \citet{franks2019flexible} used a similar parameterization in Bayesian inference and pointed out that the parametrization in Definition \ref{def::sensitivity-analysis} does not impose any testable restrictions on the observed data distributions. Nevertheless, despite the attractiveness of this parametrization, the statistical theory for sensitivity analysis is still missing for the standard estimators in the canonical setting of observational studies.

\subsection{Identification and estimation under sensitivity analysis}

This subsection will present the key identification and estimation results for $\tau$, in parallel with the results under the unconfoundedness assumption. In particular, by setting $\varepsilon_1(X) = \varepsilon_0(X) = 1$ in the formulas below, we can recover the corresponding results under the unconfoundedness assumption. We will provide theory and methods for the outcome regression, the inverse propensity score weighting, and the doubly robust estimators for $\tau$ under Definition \ref{def::sensitivity-analysis}. 

\begin{theorem}[outcome regression]
\label{thm::outcome-reg-sensitivity}
Under Definition \ref{def::sensitivity-analysis},  we have 
\begin{eqnarray}
E\{ Y(1)\mid Z=0 \}   &=& E\left\{\mu_1(X)   / \varepsilon_1(X)  \mid Z=0\right\},  \label{eq::counterfactual-1} \\
E\{ Y(0)\mid Z=1 \}   &=& E\left\{\mu_0(X)    \varepsilon_0(X)  \mid Z=1\right\}   \label{eq::counterfactual-0}
\end{eqnarray}
and therefore
\begin{eqnarray}
\tau & = & E\{ ZY + (1-Z) \mu_1(X)   / \varepsilon_1(X)\}  - E\{ Z \mu_0(X)    \varepsilon_0(X) + (1-Z)Y\}  \label{eq::predictive}\\
& = & E\{ Z \mu_1(X) + (1-Z) \mu_1(X)   / \varepsilon_1(X)\}  - E\{ Z \mu_0(X)    \varepsilon_0(X) + (1-Z) \mu_0(X)\}. \label{eq::projective}
\end{eqnarray}
\end{theorem}

In Theorem \ref{thm::outcome-reg-sensitivity}, we essentially get the identification for $E\{Y(1)\}$ and $E\{Y(0)\}$ from the identification formulas for the two counterfactual conditional means in \eqref{eq::counterfactual-1} and \eqref{eq::counterfactual-0}, which also lead to the two identification formulas for $\tau$ in \eqref{eq::predictive} and \eqref{eq::projective}. Under the unconfoundedness assumption, $\varepsilon_1(X) = \varepsilon_0(X) = 1$, \eqref{eq::projective} reduces to the first identification formula in~\eqref{eqn::identification-tau}. For general $\varepsilon_z(X)$, we need to reweight the conditional means $\mu_z(X)$ in the identification of the counterfactual means $E\{Y(z)\mid Z=1-z\}$.

With the fitted outcome model, we can construct two plug-in estimators corresponding to \eqref{eq::predictive} and \eqref{eq::projective} called the \textit{predictive} and \textit{projective} estimators for $\tau$, respectively:
\begin{eqnarray*}
\hat\tau^{\textup{pred}} &=& n^{-1}\sumn \left\{  Z_i Y_i + (1-Z_i) \hat{\mu}_1(X_i)   / \varepsilon_1(X_i)  \right\} - n^{-1}  \sumn \left\{  Z_i \hat{\mu}_0(X_i)    \varepsilon_0(X_i) + (1-Z_i) Y_i \right\},
\end{eqnarray*}
and
\begin{eqnarray*}
\hat{\tau}^{\textup{proj}}&=&n^{-1}  \sumn \left\{Z_{i}\hat{\mu}_{1}(X_{i})+\frac{(1-Z_{i})\hat{\mu}_{1}(X_{i})}{\varepsilon_{1}(X_{i})}\right\} -n^{-1}  \sumn\left\{ Z_{i}\hat{\mu}_{0}(X_{i})\varepsilon_{0}(X_{i})+(1-Z_{i})\hat{\mu}_{0}(X_{i})\right\}.
\end{eqnarray*} 
The terminologies ``predictive'' and ``projective'' are from the survey sampling literature \citep{firth1998robust, ding2018causal}. The estimators $\hat\tau^{\textup{pred}} $ and $\hat{\tau}^{\textup{proj}}$ differ slightly: the former uses the observed outcomes when available; in contrast, the latter replaces the observed outcomes with the fitted values.

More interestingly, we can also identify $\tau$ by an inverse propensity score weighting formula although Definition \ref{def::sensitivity-analysis} only involves the conditional means of the potential outcomes. 

\begin{theorem}
[inverse propensity score weighting]
\label{thm::ipw}
Under Definition \ref{def::sensitivity-analysis}, we  have 
\begin{eqnarray}\label{eq::ipw}
E\{ Y(1) \} = 
E\left\{ w_1(X) \frac{Z}{e(X)}  Y  \right\}  ,\quad
 E\{Y(0)\} = E\left\{ w_0(X) \frac{1-Z}{1-e(X)} Y \right\},
\end{eqnarray}
where
$$
w_1(X) = e(X) + \{1-e(X)\}/ \varepsilon_1(X) ,\quad
w_0(X) = e(X)  \varepsilon_0(X) + 1- e(X). 
$$
\end{theorem}

Theorem \ref{thm::ipw} modifies the classic inverse probability weighting formulas with two extra factors $w_1(X) $ and $w_0(X) $ depending on both the propensity score and the sensitivity parameters. With the fitted propensity score model, \eqref{eq::ipw} motivates the following estimator for $\tau$: 
\begin{eqnarray*} 
\hat\tau^{\textup{ht}} =  n^{-1} \sumn \hat{w}_1(X_i)\frac{Z_i Y_i}{\hat{e}(X_i) } 
- n^{-1} \sumn \hat{w}_0(X_i) \frac{ (1-Z_i) Y_i }{ 1- \hat e(X_i)},
\end{eqnarray*}
where $\hat{w}_{1}(X_{i})=\hat{e}(X_{i})+ \{1-\hat{e}(X_{i})\} / \varepsilon_{1}(X_{i})$ and $\hat{w}_{0}(X_{i})=\hat{e}(X_{i})\varepsilon_{0}(X_{i})+1-\hat{e}(X_{i})$ are the estimated weights.

The identification formulas based on outcome regression and inverse propensity score weighting motivate us to develop a combined strategy that leads to doubly robust estimation. Following \citet{bang2005doubly}, we calculate the efficient influence function for $\tau$. The details of the efficient influence function are in \cite{bickel1993efficient}, and the proof of Theorem~\ref{thm::eif} is in Section~\ref{subsec::proof_eif} in the supplementary material. Readers focusing only on applying the method do not need to understand the efficient influence function to understand our proposed estimation procedure.

\begin{theorem}[efficient influence functions]
\label{thm::eif}
Under Definition \ref{def::sensitivity-analysis}, the efficient influence function for $E\{ Y(1)\}$ is
$$
\phi_1 = w_1(X) \frac{Z}{e(X)}  Y - \frac{  \{Z-e(X)\}  \mu_1(X) }{  e(X) \varepsilon_1(X)  } - E\{ Y(1)\},
$$
the efficient influence function for $E\{ Y(0)\}$ is
$$
\phi_0 = w_0(X) \frac{1-Z}{1-e(X)}  Y - \frac{  \{e(X) - Z\}  \mu_0(X)\varepsilon_0(X)  }{ 1-   e(X) } - E\{ Y(0)\},
$$
so the efficient  influence function for $\tau $ is $\phi_1 - \phi_0.$
\end{theorem}

The efficient influence function for $\tau$ has mean $0$ by definition, so $\tau$ has the following representation: 
$$
\tau = E\left[   w_1(X) \frac{Z}{e(X)}  Y - \frac{  \{Z-e(X)\}  \mu_1(X) }{  e(X) \varepsilon_1(X)  }   \right]
-
E\left[  w_0(X) \frac{1-Z}{1-e(X)}  Y - \frac{  \{e(X) - Z\}  \mu_0(X)\varepsilon_0(X)  }{ 1-   e(X) } \right] 
$$
which motivates the following estimator based on the fitted propensity score and outcome models:
$$
\hat{\tau}^{\textup{dr}} = n^{-1} \sumn \left\{ \hat{w}_1(X_i)\frac{Z_i Y_i}{\hat{e}(X_i) } - \frac{  \check{Z}_i \hat  \mu_1(X_i) }{  \hat e(X_i) \varepsilon_1(X_i)  } - \hat{w}_0(X_i) \frac{ (1-Z_i) Y_i }{ 1- \hat e(X_i)} - \frac{ \check{Z}_i \hat  \mu_0(X_i)\varepsilon_0(X_i)  }{ 1-  \hat  e(X_i) }  \right\} .
$$ 
It is numerically identical to the following forms
\begin{eqnarray*}
\hat{\tau}^{\textup{dr}} &=& \hat\tau^{\textup{ht}}  
- n^{-1} \sumn \check Z_i  \left\{   \frac{  \hat  \mu_1(X_i) }{  \hat e(X_i) \varepsilon_1(X_i)  }    + \frac{  \hat  \mu_0(X_i)\varepsilon_0(X_i)  }{ 1-  \hat  e(X_i) } \right\} \\
&=&\hat{\tau}^{\textup{pred}}+n^{-1}\sumn\left\{ \frac{1-\hat{e}(X_{i})}{\varepsilon_{1}(X_{i})}\frac{Z_{i} \check Y_i }{\hat{e}(X_{i})}-\hat{e}(X_{i})\varepsilon_{0}(X_{i})\frac{(1-Z_{i}) \check Y_i  }{1-\hat{e}(X_{i})}\right\} \\
&=&\hat{\tau}^{\textup{proj}}+n^{-1}\sumn\left\{ \hat{w}_{1}(X_{i})\frac{Z_{i} \check Y_i  }{\hat{e}(X_{i})}-\hat{w}_{0}(X_{i})\frac{(1-Z_{i}) \check Y_i  }{1-\hat{e}(X_{i})}\right\} , 
\end{eqnarray*}
which, similar to~\eqref{eqn::doubly-robust-two-forms},  augments $\hat\tau^{\textup{ht}} $ by imputed outcomes and modifies  $\hat{\tau}^{\textup{pred}}$ and $\hat{\tau}^{\textup{proj}}$ by inverse propensity score weighted residuals.
Importantly, Theorem \ref{thm::dr} below shows that $\hat{\tau}^{\textup{dr}} $ is doubly robust.

\begin{theorem}
[double robustness]
\label{thm::dr}
Under Definition \ref{def::sensitivity-analysis}, the estimator $\hat{\tau}^{\textup{dr}} $ is consistent if either $\hat e(X)$ is consistent to $e(X)$ or $\hat\mu_z(X)$ is consistent to $\mu_z(X)$, for $z=0,1$.
\end{theorem}

Under standard regularity conditions, the doubly robust estimator $\hat{\tau}^{\textup{dr}}$ is asymptotically normally distributed with mean $\tau$ and variance achieving the variance of the efficient influence function. Therefore, we can conduct statistical inference based on the closed-form plug-in variance estimator or the nonparametric bootstrap variance estimator of $\hat{\tau}^{\textup{dr}}$. We can also construct double machine learning estimators and apply the cross-fitting technique to achieve the same asymptotic distribution under a relaxed condition on the complexity of the spaces of the nuisance parameters \citep{chernozhukov2018double}. 

\section{Calibration and specification of the sensitivity parameters}
\label{sec::calibration}

\subsection{Calibration of the sensitivity parameters}\label{subsec::calibration}

Specifying the ranges of the sensitivity parameters is fundamentally challenging in causal inference because the observed data do not provide direct information on the sensitivity parameters. We provide a method to calibrate the sensitivity parameters using the observed covariates. The identification and estimation procedures rely on the unidentifiable sensitivity parameters $(\varepsilon_1(X), \varepsilon_0(X))$. Calibration based on the observed covariates provides us guidance on their sizes, helping us to understand their relative magnitudes in the analytical framework. \cite{zhang2020calibrated} proposed a calibration method based on the observed covariates as a benchmark to understand how sensitive the estimates are with respect to an unobserved confounder in matched observational studies. In our sensitivity analysis framework, we do not directly model the unobserved confounders or their association with the treatment or potential outcome. Also, our proposed sensitivity parameter depends on the set of observed covariates $X$ we have controlled for, thus it is challenging to directly characterize the corresponding calibration of the observed covariates. Below we provide a leave-one-covariate-out approach to calibrate the sensitivity parameters. Similar approaches appeared in other sensitivity analysis methods \citep[Chapter 17 in][]{ding2023first}.

For each observed covariate $X_j$ in the $p$-dimensional $X=(X_1, \ldots, X_p)$, $j=1,\ldots,p$, we first drop it as if it is an unobserved confounder, then estimate the value of sensitivity parameters with $X_j$ unobserved. Under the unconfoundedness Assumption~\ref{assume::unconfoundedness}, we have
\begin{eqnarray*}
    \varepsilon_z(X_{-j}) &=& \frac{E\{Y(z)\mid Z=1,X_{-j}\}}{E\{Y(z)\mid Z=0,X_{-j}\}} \ =\ \frac{E\{\mu_z(X)\mid Z=1,X_{-j}\}}{E\{\mu_z(X)\mid Z=0,X_{-j}\}} 
\end{eqnarray*}
for $z=0,1$, where $X_{-j}$ denotes the $p-1$ dimensional observed covariates after dropping covariate $X_j$. The $\varepsilon_z(X_{-j})$ measures how much the observed data deviates from the unconfoundedness assumption when we delete an observed confounder $X_j$, and can be estimated using observed data. Thus, we can summarize the distribution of $\varepsilon_z(X_{-j})$ to characterize how strong a contribution each covariate $X_j$ has as a confounder. To summarize $\varepsilon_z(X_{-j})$ from a function of $X_{-j}$ to a scalar, we can further marginalize over the distribution of $X_{-j}$ to compute the mean of sensitivity parameter ignoring $X_j$, or use other summary statistics such as maximum, minimum, upper or lower quantiles of the estimated distribution to guide us on the interpretation of the magnitude of the sensitivity parameter $\varepsilon_z(X)$. We will illustrate this calibration method in Section~\ref{subsec::application}. In addition, we can also calibrate using a subset of covariates and provide a leave-multiple-covariates-out approach. The definition, identification, and estimation of the sensitivity parameters $\varepsilon_z(X_{-\mathcal{S}})$ similarly follow from the unconfoundedness assumption, where $\mathcal{S}\subset \{1,\ldots,p\}$ denotes the indices of the subset of covariates we use for calibration.

\subsection{Practical suggestions for specifying the sensitivity parameters}

In this subsection, we provide practical suggestions on specifying the sensitivity parameters and reporting the results. When the sensitivity parameters are constants not depending on $X$, i.e., $\varepsilon_1(X)=\varepsilon_1$ and $\varepsilon_0(X)=\varepsilon_0$, the proposed estimators of $\tau$ monotonically decrease in both $\varepsilon_1$ and $\varepsilon_0$. Therefore, it often suffices to present one side of the parameters for sensitivity analysis if we are interested in how strong the unobserved confounding should be to explain away the estimated causal effect under the unconfoundedness assumption. For instance, if the estimated causal effect under unconfoundedness is positive, researchers can report a table of estimated average causal effects for different $(\varepsilon_1, \varepsilon_0)$ with $\varepsilon_1 \geq 1$ and $\varepsilon_0 \geq 1$. Such table shows how robust the positive result is when we increase the sensitivity parameters, and presents how large the sensitivity parameters should be to overturn the estimated significant positive effect, while the other direction with small sensitivity parameters is usually of less interest since it will only increase the estimated positive effect. For a similar reason, when the estimated average causal effect is negative under the unconfoundedness assumption, we suggest reporting results with $\varepsilon_1 \leq 1$ and $\varepsilon_0 \leq 1$. However, in the case when we would like to learn how strong the unobserved confounding is needed to match some previously estimated causal effects with larger magnitude, the other direction may be of interest as well.

In our proposed framework, we allow for the dependence of the sensitivity parameters on observed covariates. When the sensitivity parameters depend on $X$, two alternative approaches can be pursued: modeling the sensitivity parameters explicitly or adopting a worst-case interpretation. In the first approach, we can assume $\varepsilon_z(X) = \exp(\alpha_z + \beta_z^{\T}X)$ for $z=0,1$, and specify the value of $\varepsilon_z(X)$ based on domain knowledge about the parameters $(\alpha_z,\beta_z)$. In the second approach, under the special scenarios when the outcome is non-negative (e.g., binary), treating the estimated results with constant values $(\varepsilon_1,\varepsilon_0)$ serves as a worst-case bound for the true average causal effect. Here, $\varepsilon_z$ is determined as either $\min(\varepsilon_z(X))$ or $\max(\varepsilon_z(X))$, depending on the direction of the average causal effect. We give a formal result below.

\begin{proposition}
    Let $\varepsilon_{z,\textsc{l}}$ and $\varepsilon_{z,\textsc{u}}$ denote the lower and upper bound of $\varepsilon_z(X)$, i.e., $\varepsilon_z(X)\in[\varepsilon_{z,\textsc{l}}, \varepsilon_{z,\textsc{u}}]$ for $z=0,1$. In the special case when the potential outcomes are non-negative (e.g., binary), we have $\tau \in  [\tau_\textsc{l}, \tau_\textsc{u}]$, where
    \begin{eqnarray*}
        \tau_\textsc{l} &=& E\left\{Z\mu_1(X)+\frac{(1-Z)\mu_1(X)}{\varepsilon_{1,\textsc{u}}}\right\} - E\left\{Z\mu_0(X)\varepsilon_{0,\textsc{u}} + (1-Z)\mu_0(X)\right\} \\
        &=& E\left[\left\{e(X) + \frac{1-e(X)}{\varepsilon_{1,\textsc{u}}}\right\}\frac{ZY}{e(X)}\right] - E\left[\left\{e(X)\varepsilon_{0,\textsc{u}} + 1 - e(X)\right\}\frac{(1-Z)Y}{1-e(X)}\right] \\
        &=& E\left[\left\{e(X) + \frac{1-e(X)}{\varepsilon_{1,\textsc{u}}}\right\}\frac{ZY}{e(X)} - \frac{\left\{Z-e(X)\right\}\mu_1(X)}{e(X)\varepsilon_{1,\textsc{u}}}\right] \\
        &&- E\left[\left\{e(X)\varepsilon_{0,\textsc{u}}+1-e(X)\right\}\frac{(1-Z)Y}{1-e(X)}-\frac{\left\{e(X)-Z\right\}\mu_0(X)\varepsilon_{0,\textsc{u}}}{1-e(X)}\right],
    \end{eqnarray*}
    and $\tau_\textsc{u}$ is computed using the same formulas as $\tau_\textsc{l}$, with the replacement of $\varepsilon_{1,\textsc{u}}$ and $\varepsilon_{0,\textsc{u}}$ by $\varepsilon_{1,\textsc{l}}$ and $\varepsilon_{0,\textsc{l}}$, respectively. 
\label{prop::bound_constant_epsilons}
\end{proposition}

Proposition~\ref{prop::bound_constant_epsilons} gives bounds based on the three identification formulas in Theorems~\ref{thm::outcome-reg-sensitivity}--\ref{thm::eif}. Define $\hat\tau^{*}_{\textsc{l}}$ and $\hat\tau^{*}_{\textsc{u}}$ as estimators using the same formula as $\hat \tau^{*}$ by replacing $(\varepsilon_1(X_i),\varepsilon_0(X_i))$ by $(\varepsilon_{1,\textsc{u}},\varepsilon_{0,\textsc{u}})$ and $(\varepsilon_{1,\textsc{l}},\varepsilon_{0,\textsc{l}})$, respectively, for $*=\textup{pred, proj, ht, dr}$. Following from Proposition~\ref{prop::bound_constant_epsilons}, we can treat $\hat\tau^{*}_{\textsc{l}}$ and $\hat\tau^{*}_{\textsc{u}}$ as estimators of lower and upper bounds of $\tau$, respectively. When the potential outcomes are non-positive, we have analogous results to Proposition~\ref{prop::bound_constant_epsilons}.


Suppose our estimated average causal effect under the unconfoundedness assumption is positive, we focus on $\varepsilon_1(X)\geq1$, $\varepsilon_0(X)\geq1$ to answer the question of how large the sensitivity parameters have to be to explain away the estimated positive causal effect. Looping over the pre-specified lists of constant $\varepsilon_1$ and $\varepsilon_0$, we are finding how large the lower bounds of $\varepsilon_1(X)$ and $\varepsilon_0(X)$ are to explain away the positive effect.

\section{Application and simulation study}
\label{sec::application_simulation}

\subsection{Application}\label{subsec::application}

We apply the proposed sensitivity analysis method to a real-world application. We re-analyze the observational study in \citet{bazzano2003relationship} to study whether cigarette smoking has a causal effect on homocysteine levels, the elevation of which is considered a risk factor for cardiovascular disease. We compare the homocysteine levels in daily smokers and non-smokers based on the NHANES 2005--2006 data. The NHANES is a series of surveys whose target population is the civilian, noninstitutionalized U.S. population. Since 1999, the NHANES has been conducted annually by interviewing individuals in their homes and asking the responders to complete the health examination competent of the survey. We use the cross-sectional data collected in the NHANES 2005--2006, which includes observed covariates such as gender, age, education level, BMI, and a measure of poverty. Consider the daily smokers to be in the treated group and the never smokers to be in the control group. The baseline homocysteine level in the control group is 7.86 umol/L. Under the unconfoundedness assumption with $\varepsilon_1(X)=\varepsilon_0(X)=1$, the estimated $\tau$ using the doubly robust estimator is 1.48 with a 95\% confidence interval (0.78, 2.18). We use the logit model for the propensity score and the linear model for the outcome, respectively, and use the nonparametric bootstrap method to estimate the variance. Despite a large collection of observed covariates collected in the survey, unmeasured factors such as the lifestyle of their spouses or parents, work intensity, stress level, and certain gene expressions possibly contribute to both a higher probability of smoking and higher homocysteine level. With the concern about the presence of unmeasured confounders, we conduct sensitivity analysis using the doubly robust estimator $\hat{\tau}^{\textup{dr}}$. In this example, it is challenging to specify a correct model of sensitivity parameters as a function of $X$, and the outcome of interest is non-negative. Thus, we adopt the conservative worst-case interpretation in Proposition~\ref{prop::bound_constant_epsilons}, assume the sensitivity parameters $\varepsilon_1(X)$ and $\varepsilon_0(X)$ do not depend on $X$, and vary both of them from 0.75 to 1.25. Table \ref{tab::homoscysteine} reports the results of the doubly robust estimator and its 95\% confidence interval for different $(\varepsilon_1,\varepsilon_0)$ where we again use nonparametric bootstrap to estimate the variance of $\hat{\tau}^{\textup{dr}}$. We omit the detailed results for small $\varepsilon_1$ and $\varepsilon_0$ since they all give significantly positive results at the 0.05 level. As $\varepsilon_1$ and $\varepsilon_0$ increase, the significance level decreases, and the direction of the point estimator changes when both are very large. 

However, the length of the confidence interval is not a monotone function of $\varepsilon_1(X)$ and $\varepsilon_0(X)$ since the asymptotic variance of $\hat\tau^{\textup{dr}}$ under our sensitivity analysis framework is a complicated function of the sensitivity parameters. In Section~\ref{subsec::semi_efficiency_bound} in the supplementary material, we give the specific form of the semiparametric efficiency bound for $\tau$ based on its efficient influence function in Theorem~\ref{thm::eif}.

\begin{figure}
    \centering
    \includegraphics[width=12cm]{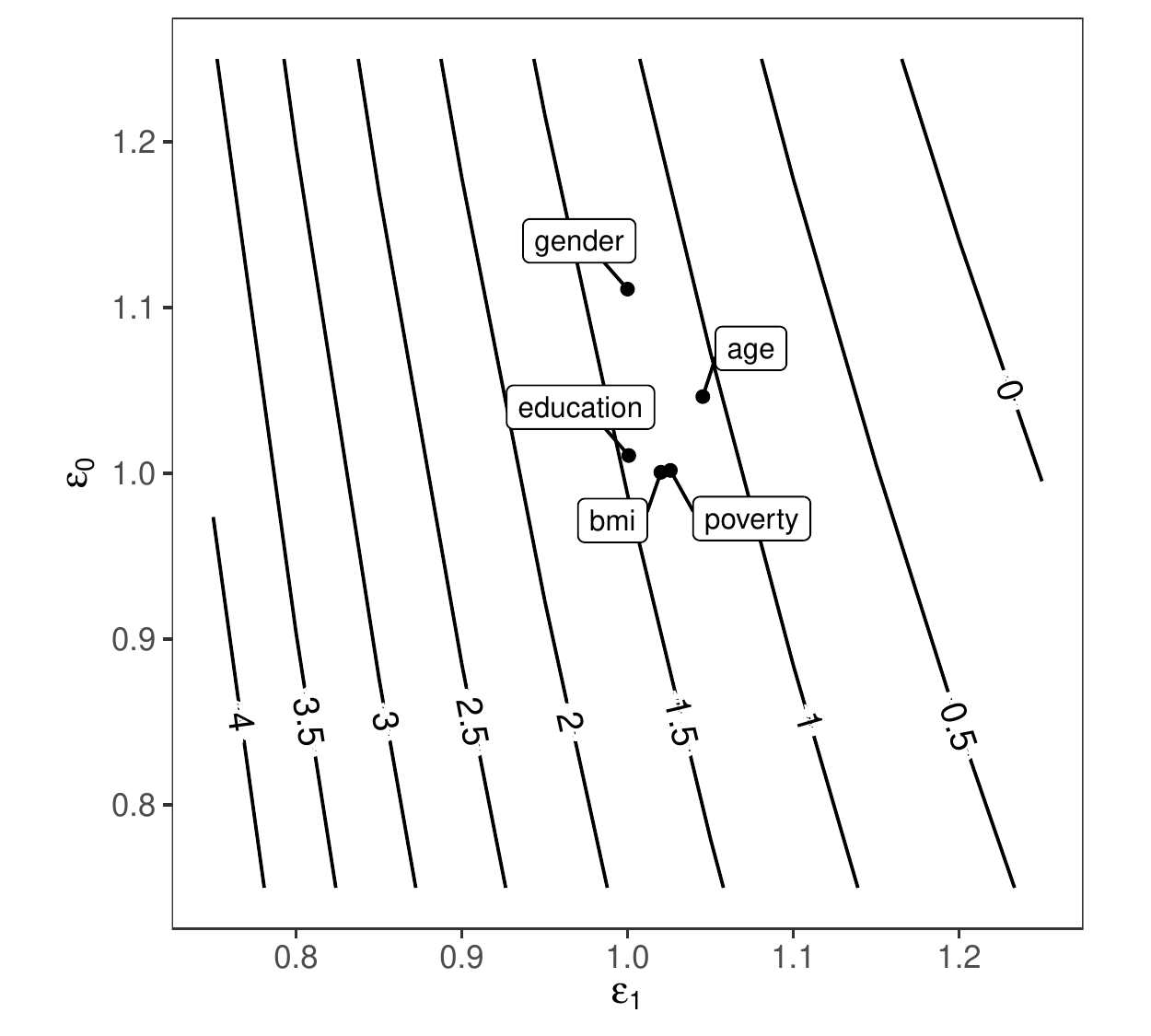}
    \caption{Sensitivity analysis for $\tau$ in the homocysteine observational study with calibration of the sensitivity parameters}
    \label{fig::contour_plot}
\end{figure}

Figure \ref{fig::contour_plot} shows the contour plot of the estimated average causal effect using the doubly robust estimator $\hat{\tau}^{\textup{dr}}$ with various values of $\varepsilon_1$ and $\varepsilon_0$, where the contour lines are generated through grid search.
We also implement the leave-one-covariate-out calibration method proposed in Section~\ref{subsec::calibration} in the example. For each observed covariate $X_j$, we label the maximum estimated value of $\varepsilon_1(X_{-j})$ and $\varepsilon_0(X_{-j})$ in the contour plot. This suggests that to explain away the estimated positive causal effect, the unobserved confounder has to be stronger than all observed covariates. Figure~\ref{fig::contour_plot} also shows that under Assumption~\ref{assume::unconfoundedness}, covariates such as BMI and poverty are weak confounders with $\varepsilon_1(X)$ and $\varepsilon_0(X)$ in the range of $\varepsilon_1(X_{-\textup{BMI}})\in[1.002, 1.020]$, $\varepsilon_0(X_{-\textup{BMI}})\in[0.998, 1.001]$ and $\varepsilon_1(X_{-\textup{poverty}})\in[0.972, 1.026]$, $\varepsilon_0(X_{-\textup{poverty}})\in[0.998, 1.002]$, respectively.

\begin{table}
\caption{Sensitivity analysis for $\tau$ in the homocysteine observational study}
\begin{small}
\doublespacing
\begin{tabular}{cc|lllll}
 & \multicolumn{1}{c}{} & \multicolumn{5}{c}{$\varepsilon_{1}$}\tabularnewline
 &  & 0.90 & 1.00 & 1.10 & 1.20 & 1.25\tabularnewline
\cline{2-2} \cline{3-3} \cline{4-4} \cline{5-5} \cline{6-6} \cline{7-7} 
\multirow{5}{*}{$\varepsilon_{0}$} & 0.90 & 2.48 (1.70, 3.27) & 1.65 (0.97, 2.33) & 0.98 (0.32, 1.63) & 0.41 (-0.21, 1.03) & 0.16 (-0.45, 0.77)\tabularnewline
 & 1.00 & 2.31 (1.58, 3.04) & 1.48 (0.78, 2.18) & 0.80 (0.12, 1.49) & 0.24 (-0.39, 0.87) & \textbf{-0.01 (-0.62, 0.60)}\tabularnewline
 & 1.10 & 2.14 (1.38, 2.89) & 1.31 (0.62, 2.00) & 0.63 (0.01, 1.25) & 0.07 (-0.56, 0.69) & \textbf{-0.18 (-0.76, 0.40)}\tabularnewline
 & 1.20 & 1.96 (1.13, 2.79) & 1.14 (0.42, 1.85) & 0.46 (-0.16, 1.07) & \textbf{-0.11 (-0.73, 0.52)} & \textbf{-0.36 (-0.97, 0.26)}\tabularnewline
 & 1.25 & 1.88 (1.10, 2.65) & 1.05 (0.35, 1.75) & 0.37 (-0.30, 1.04) & \textbf{-0.19 (-0.80, 0.41)} & \textbf{-0.44 (-1.06, 0.17)}
\end{tabular}
\end{small}
\label{tab::homoscysteine}
\end{table}


\subsection{Simulation study}
In this subsection, we conduct a simulation study to evaluate the finite sample performance of our proposed sensitivity analysis framework with the doubly robust estimator.

We generate the data using the following data-generating process with sample size $n=500$. We first generate observed covariates $X=(X_1, X_2, X_3)^\T$ with independent $X_j \sim \mathcal{N}(0, 0.5^2)$ for $j=1,2$ and $X_3 \sim \textup{Bernoulli}(0.5)$ and a binary unobserved confounder $U\sim \textup{Bernoulli}(0.5)$. Next generate the treatment $Z\mid (X,U) \sim \textup{Bernoulli}(0.25 + 0.5U)$. Then generate the potential outcomes with log-normal conditional distribution, $\log\{Y(1)\} = X_1 + X_2 + bU + e_1$ and $\log\{Y(0)\} = X_1 + X_2 + bX_3 + e_0$, where $e_1$ and $e_0$ are independent $\mathcal{N}(0, 0.5^2)$. Under this data-generating process, the population average causal effect $\tau$ equals 0. Since $U$ is an unobserved confounder of $Y(1)$ and $Z$ conditional on $X$, the unconfoundedness assumption is violated, thus estimators constructed using merely observed variables will be asymptotically biased, leading to invalid inference. Under the log-normal model of the potential outcomes and the Bernoulli treatment assignment, the true $\varepsilon_1(X)=\{0.75\exp(b) + 0.25\} / \{0.25\exp(b) + 0.75\}$, and $\varepsilon_0(X) = 1$ because the distribution of $\log\{Y(0)\}$ does not depend on $U$. We evaluate the performance of $\hat \tau^{\textup{dr}}$ with various values of $b \in (0, 0.2, 0.3, 0.5, 1, 1.5)$. For $b>0$, the true $\varepsilon_1(X) > 1$ and assuming the unconfoundedness assumption will over-estimate $\tau$. Thus, we also take a pre-specified list of values larger than 1, $\varepsilon_1(X) \in \{1, 1.10, 1.16, 1.28, 1.60, 1.93\}$ and $\varepsilon_0(X) = 1$. We use a nonparametric bootstrap with 200 bootstrap samples to estimate the asymptotic variance. For each value of $b$ and each pair $(\varepsilon_1(X),\varepsilon_0(X))$, we compute the coverage probability of the 95\% confidence interval. 

When we observe a positive $\hat \tau^{\textup{dr}}$ under the unconfoundedness assumption, at each $\varepsilon_1$ level, we reject the null hypothesis that $\tau=0$ if the constructed lower bound of the 95\% confidence interval is positive. With a positive $b$, $\tau$ will be overestimated if $\varepsilon_1$ takes a value smaller than the truth, thus it is likely to make a false rejection. While for $\varepsilon_1$ larger than the truth, we will make a conservative inference thus the false rejection is less likely. To evaluate this, we also compute the frequency of making a false rejection of a significantly positive average causal effect. The number of Monte Carlo samples is 500.

Table~\ref{tab::simulation_results} reports the simulation results including the coverage and false rejection rates of our proposed estimation procedure under various data-generating processes. First, when the sensitivity parameters are set to be equal to the true values, the coverage rate is correct. Second, the false rejection rates are valid if the value of $\varepsilon_1$ is specified to be larger than the truth. The proposed sensitivity analysis procedure has valid finite sample performance.

\begin{table}
\caption{Simulation results}
\begin{small}
\doublespacing
\begin{tabular}{cccccccccccccccc}
 &  & \multicolumn{6}{c}{\textbf{coverage rate}} &  & \multicolumn{6}{c}{\textbf{false rejection rate}}\tabularnewline
\cline{3-8} \cline{10-15}
 \multirow{2}{*}{$b$} & \multirow{2}{*}{true $\varepsilon_{1}(X)$} & \multicolumn{6}{c}{$\varepsilon_{1}$} &  & \multicolumn{6}{c}{$\varepsilon_{1}$}\tabularnewline
 &  & 1.00 & 1.10 & 1.16 & 1.28 & 1.60 & 1.93 &  & 1.00 & 1.10 & 1.16 & 1.28 & 1.60 & 1.93\tabularnewline
\hline 
\hline 
0.00 & 1.00 & 0.95 & 0.53 & 0.21 & 0.01 & 0.00 & 0.00 &  & 0.03 & 0.00 & 0.00 & 0.00 & 0.00 & 0.00\tabularnewline
0.20 & 1.10 & 0.55 & 0.96 & 0.88 & 0.32 & 0.00 & 0.00 &  & 0.45 & 0.02 & 0.00 & 0.00 & 0.00 & 0.00\tabularnewline
0.30 & 1.16 & 0.25 & 0.85 & 0.97 & 0.70 & 0.00 & 0.00 &  & 0.75 & 0.15 & 0.02 & 0.00 & 0.00 & 0.00\tabularnewline
0.50 & 1.28 & 0.02 & 0.42 & 0.71 & 0.96 & 0.19 & 0.00 &  & 0.98 & 0.58 & 0.29 & 0.02 & 0.00 & 0.00\tabularnewline
1.00 & 1.60 & 0.00 & 0.03 & 0.09 & 0.40 & 0.96 & 0.68 &  & 1.00 & 0.97 & 0.91 & 0.60 & 0.02 & 0.00\tabularnewline
1.50 & 1.93 & 0.00 & 0.01 & 0.02 & 0.11 & 0.74 & 0.95 &  & 1.00 & 0.99 & 0.98 & 0.89 & 0.26 & 0.02\tabularnewline
\end{tabular}
\end{small}
\label{tab::simulation_results}
\end{table}

\section{Extensions}\label{sec::extensions}
\subsection{Sensitivity analysis on the difference scale based on counterfactual means}
We expressed the sensitivity parameters as ratios between the observed outcome means and the corresponding counterfactual outcome means. Alternatively, we can also express the sensitivity parameters on the difference scale as Definition~\ref{def::sensitivity-analysis-difference} below \citep{robins1999association}. 

\begin{definition}
[sensitivity parameters on the difference scale]
\label{def::sensitivity-analysis-difference}
Define
\begin{eqnarray*}
 E\{ Y(1)\mid Z=1, X \}  -  E\{ Y(1)\mid Z=0, X \}  &=& \delta_1(X), \\
 E\{ Y(0)\mid Z=1, X \}  -  E\{ Y(0)\mid Z=0, X \}  &=& \delta_0(X) . 
\end{eqnarray*}
\end{definition}
In Definition~\ref{def::sensitivity-analysis-difference}, $ \delta_1(X)$ and  $\delta_0(X) $ are the two sensitivity parameters. 
 
Under Definition~\ref{def::sensitivity-analysis-difference}, we can identify the two counterfactual means by $E\{ Y(1)\mid Z=0 \} = E\left\{\mu_1(X)    -  \delta_1(X)  \mid Z=0\right\}$ and $E\{ Y(0)\mid Z=1 \} = E\left\{\mu_0(X)  +  \delta_0(X)  \mid Z=1\right\}$.
Therefore, we have the following identification formulas for $E\{ Y(1) \}$ and $E\{ Y(0) \}$ based on outcome regression, inverse probability weighting, and doubly robust estimation:
\begin{eqnarray*}\label{eq::identification_diff_scale}
    E\{ Y(1) \} &=& E \{\mu_1(X) \}   -   E\{   (1-Z)   \delta_1(X)\} \ =\ E\left\{ \frac{ZY}{e(X)} - (1-Z)   \delta_1(X) \right\} \\
    &=& E\left[ \mu_1(X) + \frac{Z\{Y-\mu_1(X)\}}{e(X)} -(1-Z)   \delta_1(X)  \right], 
\end{eqnarray*}    
and
\begin{eqnarray*}    
    E\{ Y(0) \} &=& E \{\mu_0(X) \}   +   E\{ Z   \delta_0(X)\} \ =\  E\left\{ \frac{(1-Z)Y}{1-e(X)} + Z\delta_0(X) \right\} \\
    &=& E\left[ \mu_0(X) + \frac{(1-Z)\{Y-\mu_1(X)\}}{1-e(X)} + Z\delta_0(X)  \right],
\end{eqnarray*}
where the two doubly robust formulas come from the efficient influence functions for $E\{ Y(1) \}$ and $E\{ Y(0) \}$, respectively. Taking the difference between these two expectations, we have the identification formulas for the average causal effect $\tau$, which can be written as the standard identification formulas under unconfoundedness assumption minus
$
\delta = E\{ Z  \delta_0(X)  \} + E\{   (1-Z)   \delta_1(X)\}.
$
When the sensitivity parameters $\delta_0(X)$ and $ \delta_1(X)$ do not depend on $X$, the correction simplifies to $\delta = e \delta_0 + (1-e)\delta_1$, a constant that depends on the probability of the treatment and $(\delta_1, \delta_0)$.

Similarly, for the average causal effect on the treated, the correction is
$
\delta_\textsc{t} = E\{ Z  \delta_0(X)  \} /e. 
$
When the sensitivity parameter 
$\delta_0(X)$ does not depend on $X$, the correction simplifies to $\delta_0$, a constant that equals the sensitivity parameter itself.

From the above discussion, sensitivity analysis based on the difference scale is much simpler than that based on the ratio scale. Its simplicity is an advantage. However, it also has disadvantages. First, the sensitivity analysis estimators reduce to the corresponding classic estimators minus prespecified constants. It makes sensitivity analysis a tautology of specifying how different the estimators and the estimands are. 
Second, the sensitivity parameters in Definition \ref{def::sensitivity-analysis-difference} depend on the baseline expectations of the outcome. They may be harder to specify than those in Definition \ref{def::sensitivity-analysis}. This is a classic reason to focus on the ratio scale in sensitivity analysis \citep{cornfield1959smoking, poole2010origin, ding2014generalized}. 

\citet{sjolander2022sensitivity} generalized the sensitivity parameters on the difference scale to allow for the transformation of the conditional means by a smooth link function and proposed an outcome regression approach to conduct sensitivity analyses with generalized linear models.

\subsection{Extension to nonlinear causal parameters}
We can extend the current sensitivity analysis framework to analyze a more general class of nonlinear causal parameters, $g(\mu_1,\mu_0)$, a general function of the marginal means of the potential outcomes. Previous results focus on the special case when $g(\mu_1,\mu_0)=\mu_1-\mu_0$. More generally, many other function forms are of interest. Specifically, when the potential outcomes are binary, the causal risk ratio and the causal odds ratio,
$\textsc{rr} = \mu_1 / \mu_0$ and $\textsc{or} = \{\mu_1 / (1-\mu_1)\} / \{\mu_0 / (1-\mu_0)\}$
are two canonical causal parameters. We can utilize the nonparametric identification and proposed estimators for $\mu_1$ and $\mu_0$ and construct estimators for a general causal parameter by simply plugging in estimators $\hat\mu_z^{*}$ into function $g(\mu_1,\mu_0)$ for $*\in\{\textup{pred, prod, ht, dr}\}$ and $z=0,1$. The plug-in outcome and inverse probability weighting estimators are straightforward. The plug-in estimator $g(\hat\mu_1^{\textup{dr}}, \hat\mu_0^{\textup{dr}})$ remains to be doubly robust and achieves the semiparametric efficiency bound. We also implement estimators for $\textsc{rr}$, $\textsc{or}$, and their logarithms in the R package \texttt{saci}.

\subsection{Other extensions in the supplementary material}

For the inverse propensity score weighting and doubly robust estimators, we presented only the Horvitz--Thompson-type estimators. It is straightforward to obtain the corresponding Hajek-type estimators \citep{hajek1971comment}. We present them in Section~\ref{subsec::hajek} in the supplementary material.

Another canonical estimation strategy is matching \citep{rubin2006matched}. \cite{lin2021estimation} pointed out that \cite{abadie2011bias}'s bias-corrected matching estimator has an equivalent form as the doubly robust estimator if we view matching as a nonparametric method to estimate the propensity score. Therefore, similar sensitivity analysis formulas also hold for matching. We present them in  Section~\ref{subsec::matching} in the supplementary material.

Another parameter of interest is the average causal effect on the treated units. Under our sensitivity analysis framework, we provide identification results, estimation procedures, and real-world examples to estimate the average causal effect on the treated units. We also implement estimators in our R package. We present details in Section~\ref{subsec::att} in the supplementary material. 

We also extend our framework of sensitivity analysis to survival outcomes with the parameter of interest defined as the difference between two survival functions $\tau(t) = \pr\{Y(1)>t\} - \pr\{Y(0)>t\}$ and to observational studies with a multi-valued treatment. Moreover, estimating the controlled direct effect reduces to estimating causal effects with a multi-valued treatment if we view the treatment and the intermediate variable as two treatment factors \citep{vanderweele2015explanation}. Again, we relegate the details to Sections~\ref{subsec::survival} and~\ref{subsec::multi-valued} in the supplementary material.

\section{Discussion}
\label{sec::discussion}
We focus on the cross-sectional setting, while the framework can be extended to deal with longitudinal data \citep{robins1999association}. With a time-varying treatment, the sensitivity parameters are also time-varying and depend on the past treatment trajectory. We can derive similar nonparametric identification formulas by modifying the g-formula or the inverse probability weighting estimators under sequential ignorability. We leave the derivation of the semiparametric efficient influence function and the implementation of the motivated estimator to future research.

\section*{Acknowledgements}

Peng Ding was partially supported by the U.S. National Science Foundation \# 1945136. A reviewer and the Associate Editor made constructive comments.

\bibliographystyle{biom}
\bibliography{ssa_bib}

\begin{thebibliography}{}

\bibitem[\protect\citeauthoryear{Abadie and Imbens}{Abadie and
  Imbens}{2006}]{abadie2006large}
Abadie, A. and Imbens, G.~W. (2006).
\newblock Large sample properties of matching estimators for average treatment
  effects.
\newblock {\em Econometrica} {\bf 74,} 235--267.

\bibitem[\protect\citeauthoryear{Abadie and Imbens}{Abadie and
  Imbens}{2011}]{abadie2011bias}
Abadie, A. and Imbens, G.~W. (2011).
\newblock Bias-corrected matching estimators for average treatment effects.
\newblock {\em Journal of Business and Economic Statistics} {\bf 29,} 1--11.

\bibitem[\protect\citeauthoryear{Bang and Robins}{Bang and
  Robins}{2005}]{bang2005doubly}
Bang, H. and Robins, J.~M. (2005).
\newblock Doubly robust estimation in missing data and causal inference models.
\newblock {\em Biometrics} {\bf 61,} 962--973.

\bibitem[\protect\citeauthoryear{Bazzano, He, Muntner, Vupputuri, and
  Whelton}{Bazzano et~al.}{2003}]{bazzano2003relationship}
Bazzano, L.~A., He, J., Muntner, P., Vupputuri, S., and Whelton, P.~K. (2003).
\newblock Relationship between cigarette smoking and novel risk factors for
  cardiovascular disease in the united states.
\newblock {\em Annals of Internal Medicine} {\bf 138,} 891--897.

\bibitem[\protect\citeauthoryear{Bickel, Klaassen, Ritov, and Wellner}{Bickel
  et~al.}{1993}]{bickel1993efficient}
Bickel, P.~J., Klaassen, C.~A., Ritov, Y., and Wellner, J.~A. (1993).
\newblock {\em Efficient and Adaptive Estimation for Semiparametric Models}.
\newblock Baltimore: Johns Hopkins University Press.

\bibitem[\protect\citeauthoryear{Chernozhukov, Chetverikov, Demirer, Duflo,
  Hansen, Newey, and Robins}{Chernozhukov
  et~al.}{2018}]{chernozhukov2018double}
Chernozhukov, V., Chetverikov, D., Demirer, M., Duflo, E., Hansen, C., Newey,
  W., and Robins, J. (2018).
\newblock Double/debiased machine learning for treatment and structural
  parameters.
\newblock {\em Econometrics Journal} {\bf 21,} C1--C68.

\bibitem[\protect\citeauthoryear{Cornfield, Haenszel, Hammond, Lilienfeld,
  Shimkin, and Wynder}{Cornfield et~al.}{1959}]{cornfield1959smoking}
Cornfield, J., Haenszel, W., Hammond, E.~C., Lilienfeld, A.~M., Shimkin, M.~B.,
  and Wynder, E.~L. (1959).
\newblock Smoking and lung cancer: recent evidence and a discussion of some
  questions.
\newblock {\em Journal of the National Cancer Institute} {\bf 22,} 173--203.

\bibitem[\protect\citeauthoryear{Cox}{Cox}{1972}]{cox1972regression}
Cox, D.~R. (1972).
\newblock Regression models and life-tables.
\newblock {\em Journal of the Royal Statistical Society: Series B
  (Methodological)} {\bf 34,} 187--202.

\bibitem[\protect\citeauthoryear{Ding}{Ding}{2024}]{ding2023first}
Ding, P. (2024).
\newblock {\em A First Course in Causal Inference}.
\newblock New York: Chapman \& Hall.

\bibitem[\protect\citeauthoryear{Ding and Li}{Ding and
  Li}{2018}]{ding2018causal}
Ding, P. and Li, F. (2018).
\newblock Causal inference: a missing data perspective.
\newblock {\em Statistical Science} {\bf 33,} 214--237.

\bibitem[\protect\citeauthoryear{Ding and Lu}{Ding and
  Lu}{2017}]{ding2017principal}
Ding, P. and Lu, J. (2017).
\newblock Principal stratification analysis using principal scores.
\newblock {\em Journal of the Royal Statistical Society: Series B (Statistical
  Methodology)} {\bf 79,} 757--777.

\bibitem[\protect\citeauthoryear{Ding and VanderWeele}{Ding and
  VanderWeele}{2014}]{ding2014generalized}
Ding, P. and VanderWeele, T.~J. (2014).
\newblock {Generalized Cornfield conditions for the risk difference}.
\newblock {\em Biometrika} {\bf 101,} 971--977.

\bibitem[\protect\citeauthoryear{Ding and VanderWeele}{Ding and
  VanderWeele}{2016}]{ding2016sensitivity}
Ding, P. and VanderWeele, T.~J. (2016).
\newblock Sensitivity analysis without assumptions.
\newblock {\em Epidemiology} {\bf 27,} 368--377.

\bibitem[\protect\citeauthoryear{Dorn and Guo}{Dorn and
  Guo}{2022}]{dorn2022sharp}
Dorn, J. and Guo, K. (2022).
\newblock Sharp sensitivity analysis for inverse propensity weighting via
  quantile balancing.
\newblock {\em Journal of the American Statistical Association} {\bf 00,}
  1--13.

\bibitem[\protect\citeauthoryear{Firth and Bennett}{Firth and
  Bennett}{1998}]{firth1998robust}
Firth, D. and Bennett, K.~E. (1998).
\newblock Robust models in probability sampling (with discussion).
\newblock {\em Journal of the Royal Statistical Society: Series B (Statistical
  Methodology)} {\bf 60,} 3--21.

\bibitem[\protect\citeauthoryear{Franks, D’Amour, and Feller}{Franks
  et~al.}{2020}]{franks2019flexible}
Franks, A., D’Amour, A., and Feller, A. (2020).
\newblock Flexible sensitivity analysis for observational studies without
  observable implications.
\newblock {\em Journal of the American Statistical Association} {\bf 115,}
  1730--1746.

\bibitem[\protect\citeauthoryear{Hahn}{Hahn}{1998}]{hahn1998role}
Hahn, J. (1998).
\newblock On the role of the propensity score in efficient semiparametric
  estimation of average treatment effects.
\newblock {\em Econometrica} {\bf 66,} 315--331.

\bibitem[\protect\citeauthoryear{H{\'a}jek}{H{\'a}jek}{1971}]{hajek1971comment}
H{\'a}jek, J. (1971).
\newblock Comment: An essay on the logical foundations of survey sampling, part
  one.
\newblock {\em The Foundations of Survey Sampling} {\bf 236,}.

\bibitem[\protect\citeauthoryear{Hern{\'a}n}{Hern{\'a}n}{2010}]{hernan2010hazards}
Hern{\'a}n, M.~A. (2010).
\newblock The hazards of hazard ratios.
\newblock {\em Epidemiology} {\bf 21,} 13--15.

\bibitem[\protect\citeauthoryear{Imai and Van~Dyk}{Imai and
  Van~Dyk}{2004}]{imai2004causal}
Imai, K. and Van~Dyk, D.~A. (2004).
\newblock Causal inference with general treatment regimes: Generalizing the
  propensity score.
\newblock {\em Journal of the American Statistical Association} {\bf 99,}
  854--866.

\bibitem[\protect\citeauthoryear{Imbens}{Imbens}{2000}]{imbens2000role}
Imbens, G.~W. (2000).
\newblock The role of the propensity score in estimating dose-response
  functions.
\newblock {\em Biometrika} {\bf 87,} 706--710.

\bibitem[\protect\citeauthoryear{Imbens}{Imbens}{2003}]{imbens2003sensitivity}
Imbens, G.~W. (2003).
\newblock Sensitivity to exogeneity assumptions in program evaluation.
\newblock {\em American Economic Review} {\bf 93,} 126--132.

\bibitem[\protect\citeauthoryear{Jiang, Yang, and Ding}{Jiang
  et~al.}{2022}]{jiang2020multiply}
Jiang, Z., Yang, S., and Ding, P. (2022).
\newblock Multiply robust estimation of causal effects under principal
  ignorability.
\newblock {\em Journal of the Royal Statistical Society: Series B (Statistical
  Methodology)} {\bf 84,} 1423--1445.

\bibitem[\protect\citeauthoryear{Kasza, Wolfe, and Schuster}{Kasza
  et~al.}{2017}]{kasza2017assessing}
Kasza, J., Wolfe, R., and Schuster, T. (2017).
\newblock Assessing the impact of unmeasured confounding for binary outcomes
  using confounding functions.
\newblock {\em International Journal of Epidemiology} {\bf 46,} 1303--1311.

\bibitem[\protect\citeauthoryear{Lin, Psaty, and Kronmal}{Lin
  et~al.}{1998}]{lin1998assessing}
Lin, D.~Y., Psaty, B.~M., and Kronmal, R.~A. (1998).
\newblock Assessing the sensitivity of regression results to unmeasured
  confounders in observational studies.
\newblock {\em Biometrics} {\bf 54,} 948--963.

\bibitem[\protect\citeauthoryear{Lin, Ding, and Han}{Lin
  et~al.}{2023}]{lin2021estimation}
Lin, Z., Ding, P., and Han, F. (2023).
\newblock Estimation based on nearest neighbor matching: from density ratio to
  average treatment effect.
\newblock {\em Econometrica} {\bf 91,} 2187--2217.

\bibitem[\protect\citeauthoryear{Poole}{Poole}{2010}]{poole2010origin}
Poole, C. (2010).
\newblock On the origin of risk relativism.
\newblock {\em Epidemiology} {\bf 21,} 3--9.

\bibitem[\protect\citeauthoryear{Robins}{Robins}{1999}]{robins1999association}
Robins, J.~M. (1999).
\newblock Association, causation, and marginal structural models.
\newblock {\em Synthese} {\bf 121,} 151--179.

\bibitem[\protect\citeauthoryear{Rosenbaum}{Rosenbaum}{1987}]{rosenbaum1987sensitivity}
Rosenbaum, P.~R. (1987).
\newblock Sensitivity analysis for certain permutation inferences in matched
  observational studies.
\newblock {\em Biometrika} {\bf 74,} 13--26.

\bibitem[\protect\citeauthoryear{Rosenbaum and Rubin}{Rosenbaum and
  Rubin}{1983a}]{Rosenbaum::1983JRSSB}
Rosenbaum, P.~R. and Rubin, D.~B. (1983a).
\newblock Assessing sensitivity to an unobserved binary covariate in an
  observational study with binary outcome.
\newblock {\em Journal of the Royal Statistical Society: Series B (Statistical
  Methodology)} {\bf 45,} 212--218.

\bibitem[\protect\citeauthoryear{Rosenbaum and Rubin}{Rosenbaum and
  Rubin}{1983b}]{rosenbaum1983central}
Rosenbaum, P.~R. and Rubin, D.~B. (1983b).
\newblock The central role of the propensity score in observational studies for
  causal effects.
\newblock {\em Biometrika} {\bf 70,} 41--55.

\bibitem[\protect\citeauthoryear{Rubin}{Rubin}{1978}]{rubin1978bayesian}
Rubin, D.~B. (1978).
\newblock Bayesian inference for causal effects: The role of randomization.
\newblock {\em Annals of Statistics} {\bf 6,} 34--58.

\bibitem[\protect\citeauthoryear{Rubin}{Rubin}{2006}]{rubin2006matched}
Rubin, D.~B. (2006).
\newblock {\em Matched Sampling for Causal Effects}.
\newblock Cambridge: Cambridge University Press.

\bibitem[\protect\citeauthoryear{Scharfstein, Rotnitzky, and
  Robins}{Scharfstein et~al.}{1999}]{scharfstein1999adjusting}
Scharfstein, D.~O., Rotnitzky, A., and Robins, J.~M. (1999).
\newblock Adjusting for nonignorable drop-out using semiparametric nonresponse
  models.
\newblock {\em Journal of the American Statistical Association} {\bf 94,}
  1096--1120.

\bibitem[\protect\citeauthoryear{Sj{\"o}lander, Gabriel, and
  Cioc{\u{a}}nea-Teodorescu}{Sj{\"o}lander
  et~al.}{2022}]{sjolander2022sensitivity}
Sj{\"o}lander, A., Gabriel, E.~E., and Cioc{\u{a}}nea-Teodorescu, I. (2022).
\newblock Sensitivity analysis for causal effects with generalized linear
  models.
\newblock {\em Journal of Causal Inference} {\bf 10,} 441--479.

\bibitem[\protect\citeauthoryear{Tchetgen and Shpitser}{Tchetgen and
  Shpitser}{2012}]{tchetgen2012semiparametric}
Tchetgen, E. J.~T. and Shpitser, I. (2012).
\newblock Semiparametric theory for causal mediation analysis: efficiency
  bounds, multiple robustness, and sensitivity analysis.
\newblock {\em Annals of Statistics} {\bf 40,} 1816--1845.

\bibitem[\protect\citeauthoryear{VanderWeele}{VanderWeele}{2015}]{vanderweele2015explanation}
VanderWeele, T.~J. (2015).
\newblock {\em {Explanation in Causal Inference: Methods for Mediation and
  Interaction}}.
\newblock Oxford: Oxford University Press.

\bibitem[\protect\citeauthoryear{VanderWeele and Ding}{VanderWeele and
  Ding}{2017}]{vanderweele2017sensitivity}
VanderWeele, T.~J. and Ding, P. (2017).
\newblock Sensitivity analysis in observational research: introducing the
  e-value.
\newblock {\em Annals of Internal Medicine} {\bf 167,} 268--274.

\bibitem[\protect\citeauthoryear{VanderWeele, Tchetgen, and
  Halloran}{VanderWeele et~al.}{2014}]{vanderweele2014interference}
VanderWeele, T.~J., Tchetgen, E. J.~T., and Halloran, M.~E. (2014).
\newblock Interference and sensitivity analysis.
\newblock {\em Statistical science: a review journal of the Institute of
  Mathematical Statistics} {\bf 29,} 687.

\bibitem[\protect\citeauthoryear{Xie and Liu}{Xie and
  Liu}{2005}]{xie2005adjusted}
Xie, J. and Liu, C. (2005).
\newblock Adjusted kaplan--meier estimator and log-rank test with inverse
  probability of treatment weighting for survival data.
\newblock {\em Statistics in Medicine} {\bf 24,} 3089--3110.

\bibitem[\protect\citeauthoryear{Yang and Lok}{Yang and
  Lok}{2018}]{yang2018sensitivity}
Yang, S. and Lok, J.~J. (2018).
\newblock Sensitivity analysis for unmeasured confounding in coarse structural
  nested mean models.
\newblock {\em Statistica Sinica} {\bf 28,} 1703--1723.

\bibitem[\protect\citeauthoryear{Zhang and Small}{Zhang and
  Small}{2020}]{zhang2020calibrated}
Zhang, B. and Small, D.~S. (2020).
\newblock A calibrated sensitivity analysis for matched observational studies
  with application to the effect of second-hand smoke exposure on blood lead
  levels in children.
\newblock {\em Journal of the Royal Statistical Society Series C: Applied
  Statistics} {\bf 69,} 1285--1305.

\bibitem[\protect\citeauthoryear{Zhao, Small, and Bhattacharya}{Zhao
  et~al.}{2019}]{zhao2019sensitivity}
Zhao, Q., Small, D.~S., and Bhattacharya, B.~B. (2019).
\newblock Sensitivity analysis for inverse probability weighting estimators via
  the percentile bootstrap.
\newblock {\em Journal of the Royal Statistical Society: Series B (Statistical
  Methodology)} {\bf 81,} 735--761.

\bibitem[\protect\citeauthoryear{Zubizarreta, Cerda, and Rosenbaum}{Zubizarreta
  et~al.}{2013}]{zubizarreta2013effect}
Zubizarreta, J.~R., Cerda, M., and Rosenbaum, P.~R. (2013).
\newblock Effect of the 2010 chilean earthquake on posttraumatic stress
  reducing sensitivity to unmeasured bias through study design.
\newblock {\em Epidemiology} {\bf 24,} 79--87.

\end{thebibliography}

\newpage

\appendix

\begin{center}
  \LARGE {\bf Supplementary Material}
\end{center}

\pagenumbering{arabic} 
\renewcommand*{\thepage}{S\arabic{page}}

\setcounter{lemma}{0} 
\global\long\def\thelemma{\textup{S}\arabic{lemma}}%
 \setcounter{equation}{0} 
\global\long\def\theequation{S\arabic{equation}}%
 \setcounter{section}{0} 
\global\long\def\thesection{S\arabic{section}}%
 \setcounter{assumption}{0} 
\global\long\def\theassumption{S\arabic{assumption}}%
 \setcounter{theorem}{0} 
\global\long\def\thetheorem{S\arabic{theorem}}%
 \setcounter{proposition}{0} 
\global\long\def\theproposition{S\arabic{proposition}}%
 \setcounter{definition}{0} 
\global\long\def\thedefinition{S\arabic{definition}}%
 \setcounter{example}{0} 
\global\long\def\theexample{S\arabic{example}}%
 \setcounter{figure}{0} 
\global\long\def\thefigure{S\arabic{figure}}%
 
\def\thesubsection{S\arabic{section}.\arabic{subsection}}%

The supplementary materials contain the following sections.

Section \ref{sec::additional} presents additional results on several extensions, formulas for the Hajek-type estimators and for the bias-corrected matching estimator, semiparametric efficiency bounds for $\tau$ and $\tau_\T$, and connection of the sensitivity parameters to the treatment selection model.

Section \ref{sec::R-package} provides guidance to our R package and presents additional data analysis results.

Section \ref{sec::proofs} presents all the proofs.

\section{Additional results}\label{sec::additional}

\subsection{Extension to the average treatment effect on the treated units} \label{subsec::att}
Another parameter of interest is the average treatment effect on the treated units
$$
\tau_\textsc{t} = E\{  Y(1) - Y(0)  \mid Z=1  \} 
= E(Y\mid Z=1) - E\{  Y(0)  \mid Z=1  \} .
$$
Let $\mu_{\textsc{t}1} = E(Y\mid Z=1)$ and $\mu_{\textsc{t}0} =  E\{  Y(0)  \mid Z=1  \}$ denote these two terms. The first term $\mu_{\textsc{t}1}$ has a simple moment estimator $ \hat{\mu}_{\textsc{t}1} =  n_1^{-1}  \sumn Z_i Y_i  $, where $n_1 = \sumn Z_i$ is the total number of treated units. The key is to identify and estimate the second term $\mu_{\textsc{t}0}$. In parallel with Theorems \ref{thm::outcome-reg-sensitivity} and \ref{thm::ipw}, $\mu_{\textsc{t}0}$ has two identification forms based on the outcome and the propensity score models. Define $e = \pr(Z=1)$. 

\begin{theorem}
[outcome regression and inverse propensity score weighting for $\tau_\textsc{t} $]
\label{thm::att-reg-ipw} 
Under Definition \ref{def::sensitivity-analysis}, we have 
 \begin{eqnarray*}
 E\{  Y(0)  \mid Z=1  \}  &=&  E\left\{  Z \mu_0(X)  \varepsilon_0(X)    \right\} /e \\
 &=& E\left\{  e(X)\varepsilon_0(X) \frac{ 1-Z}{1-e(X)}  Y      \right\} /e.
\end{eqnarray*}
\end{theorem}

Theorem \ref{thm::att-reg-ipw} motivates using $\hat{\tau}_{\textsc{t}}^* = \hat{\mu}_{\textsc{t}1}  - \hat{\mu}_{\textsc{t}0}^* \ (* = \textup{reg},\textup{ht})$ to estimate $\tau_{\textsc{t}}$, where 
 \begin{eqnarray*}  
\hat{\mu}_{\textsc{t}0}^\textup{reg} &=&  n_1^{-1} \sumn Z_i  \varepsilon_0(X_i) \hat  \mu_0(X_i) ,   \\
\hat{\mu}_{\textsc{t}0}^\textup{ht} &=&  n_1^{-1} \sumn   \varepsilon_0(X_i) \hat  o(X_i) (1-Z_i) Y_i , 
\end{eqnarray*}  
with the estimated conditional odds of the treatment given covariates, $  \hat  o(X_i) = \hat  e(X_i) / \{1- \hat  e(X_i)\}$. To motivate the doubly robust estimator for $\tau_{\textsc{t}}$, we also derive the efficient influence function for $  \mu_{\textsc{t}0}  $.

\begin{theorem}
[efficient influence function for $  \mu_{\textsc{t}0}  $]
\label{thm::eif-att}
Under Definition \ref{def::sensitivity-analysis}, the efficient influence function for $  \mu_{\textsc{t}0}  $ equals
$$
\phi_{\textsc{t}0} = \left[  Z\left\{\varepsilon_0(X)\mu_0(X) - \mu_{\textsc{t}0}\right\} + \varepsilon_0(X)e(X)\frac{(1-Z)\{Y-\mu_0(X)\}}{1-e(X)} \right] / e. 
$$

\end{theorem}

The efficient influence function for $  \mu_{\textsc{t}0}  $ has mean $0$, so $  \mu_{\textsc{t}0}  $ has the following representation:
$$
  \mu_{\textsc{t}0}   = E \left[  Z \varepsilon_0(X)\mu_0(X)   + \varepsilon_0(X)e(X)\frac{(1-Z)\{Y-\mu_0(X)\}}{1-e(X)} \right] / e .
$$
This representation motivates the following estimator for $\tau_\textsc{t} $: 
%
%
 \begin{eqnarray*}  
 \hat{\mu}_{\textsc{t}0}^{\textup{dr}}
 &=&\hat{\mu}_{\textsc{t}0}^{\text{reg}}+n_{1}^{-1}\sumn\varepsilon_{0}(X_{i})\hat{o}(X_{i})(1-Z_{i}) \check{Y}_i \\ 
&=& \hat{\mu}_{\textsc{t}0}^{\textup{ht}} + n_{1}^{-1}\sumn\varepsilon_{0}(X_{i})\frac{\check{Z}_i \hat{\mu}_{0}(X_{i})}{1-\hat{e}(X_{i})} ,
\end{eqnarray*}   
which are two numerically identical forms based on the fitted propensity score and outcome models.

Importantly, Theorem \ref{thm::dr-att} below shows that $    \hat{\mu}_{\textsc{t}0}^{\textup{dr}}$ is doubly robust.
 
\begin{theorem}
[double robustness for estimating $\tau_{\textsc{t}}$]
\label{thm::dr-att}
Under Definition \ref{def::sensitivity-analysis}, the estimator $\hat{\mu}_{\textsc{t}0}^{\textup{dr}} $ is doubly robust in the sense that it is consistent if either the propensity score or the outcome model is correctly specified. 
\end{theorem}

We also provide a worst-case interpretation for the average treatment effect on the treated with a constant value of $\varepsilon_0$ in the special scenarios when the outcome is non-negative (e.g., binary). Again, $\varepsilon_0$ is determined as either $\min(\varepsilon_0(X))$ or $\max(\varepsilon_0(X))$, depending on the direction of the average treatment effect on the treated. We give a formal in the following proposition.

\begin{proposition}
    Let $\varepsilon_{0,\textsc{l}}$ and $\varepsilon_{0,\textsc{u}}$ denote the lower and upper bound of $\varepsilon_0(X)$, i.e., $\varepsilon_0(X)\in[\varepsilon_{0,\textsc{l}}, \varepsilon_{0,\textsc{u}}]$. In the special case when the potential outcomes are non-negative (e.g., binary), we have $\tau_\T \in  [\tau_{\T,\textsc{l}}, \tau_{\T,\textsc{u}}]$, where
    \begin{eqnarray*}
        \tau_{\T,\textsc{l}} &=& E(Y\mid Z=1) - \varepsilon_{0,\textsc{u}}E\left\{Z\mu_0(X)\right\}/e \\
        &=& E(Y\mid Z=1) - \varepsilon_{0,\textsc{u}}E\left[\frac{e(X)(1-Z)Y}{e\left\{1-e(X)\right\}}\right] \\
        &=& E(Y\mid Z=1) - \varepsilon_{0,\textsc{u}}E\left[Z\mu_0(X)+e(X)\frac{(1-Z)\left\{Y-\mu_0(X)\right\}}{1-e(X)}\right]/e,
    \end{eqnarray*}
    and $\tau_{\T,\textsc{u}}$ is computed using the same formula as $\tau_{\T,\textsc{l}}$, with the replacement of $\varepsilon_{0,\textsc{u}}$ by $\varepsilon_{0,\textsc{l}}$. 
\label{prop::bound_constant_epsilons_att}
\end{proposition}

Proposition~\ref{prop::bound_constant_epsilons_att} gives bounds based on the three identification formulas based on the outcome model, the propensity score model, and the efficient influence function for $\tau_\T$. Define $\hat\tau^{*}_{\T,\textsc{l}}$ and $\hat\tau^{*}_{\T,\textsc{u}}$ $\hat \tau^{*}_\T$ by replacing $\varepsilon_0(X_i)$ by $\varepsilon_{0,\textsc{u}}$ and $\varepsilon_{0,\textsc{l}}$, respectively, for $*=\textup{reg, ht, dr}$. Following from Proposition~\ref{prop::bound_constant_epsilons_att}, we can treat $\hat\tau^{*}_{\T,\textsc{l}}$ and $\hat\tau^{*}_{\T,\textsc{u}}$ as estimators of lower and upper bounds of $\tau_\T$, respectively. Symmetrically, when the potential outcomes are non-positive, we have analogous results to Proposition~\ref{prop::bound_constant_epsilons_att}.

\begin{example} \label{eg::ATT}
We continue the application study in Section~\ref{subsec::application} by applying the above results to estimate $\tau_{\textsc{t}}$ with the same observational data and focus on the other causal parameter of interest, the average treatment effect on the treated $\tau_{\textsc{t}}$. The estimated $\tau_{\textsc{t}}$ under the unconfoundedness assumption is 1.36, with a 95\% confidence interval (0.66, 2.05), using the same unit of the homocysteine level umol/L. We conduct sensitivity analysis using the doubly robust estimator $\hat{\tau}_{\textsc{t}}^{\textup{dr}}$. The estimated $\tau_{\textsc{t}}$ only depends on values of $\varepsilon_0$, we again simplify it not to depend on $X$ and vary it from 0.75 to 1.25, and report the results for larger values of $\varepsilon_0$, since for values smaller than 1, the estimated $\tau_{\textsc{t}}$ are all significantly positive at the 0.05 level. Table \ref{tab::homoscysteine_att} summarizes the results. For $\varepsilon_0 \geq 1.1$, the inference changes, and for $\varepsilon_0 \geq 1.2$, the estimated $\tau_{\textsc{t}}$ becomes negative.
\end{example}

\begin{table}
\caption{Sensitivity analysis for $\tau_{\textsc{t}}$ in the homocysteine observational study}
\begin{center}
\begin{small}
\doublespacing
\begin{tabular}{l|l}
$\varepsilon_{0}$ & \multicolumn{1}{c}{$\tau_{\textsc{t}}$}\tabularnewline
\hline 
0.90 & 2.18 (1.41, 2.96)\tabularnewline
0.95 & 1.77 (1.07, 2.46)\tabularnewline
1.00 & 1.36 (0.66, 2.05)\tabularnewline
1.05 & 0.94 (0.19, 1.70)\tabularnewline
1.10 & 0.53 (-0.22, 1.29)\tabularnewline
1.15 & 0.12 (-0.64, 0.88)\tabularnewline
1.20 & \textbf{-0.29 (-1.05, 0.47)}\tabularnewline
1.25 & \textbf{-0.70 (-1.52, 0.11)}\tabularnewline
\end{tabular}
\end{small}
\end{center}
\label{tab::homoscysteine_att}
\end{table}

\subsection{Extension to survival outcomes}
\label{subsec::survival}

We can also extend the results to survival outcomes. To avoid the problem of the hazard ratio for causal inference that it has a built-in selection bias even under randomization \citep{hernan2010hazards},  we define the parameter of interest as the difference between the two survival functions
$$
\tau(t) = S_1(t) - S_0(t),
$$ 
where $S_z(t) = \pr\{Y(z)>t\}$ denotes the potential survival functions  for $z\in\{0,1\}$. For a fixed $t$, $\tau(t)$ is the average causal effect on the indicator $1(Y>t)$. Define the sensitivity parameters as 
\allowdisplaybreaks
\begin{eqnarray*}
    \frac{ \pr\{ Y(1)>t\mid Z=1, X \} }{ \pr\{ Y(1)>t\mid Z=0, X \} } = \varepsilon_{1t}(X), \quad 
    \frac{ \pr\{ Y(0)>t\mid Z=1, X \} }{ \pr\{ Y(0)>t\mid Z=0, X \} } = \varepsilon_{0t}(X) . \label{def::epsilon_survival}
\end{eqnarray*}
Define $p_{zt}(X)=\pr(Y>t\mid Z=z, X)$ as the conditional survival probability for $z\in\{0,1\}$. 
We can rewrite the survival function of potential outcomes as
\begin{eqnarray}
    S_1(t) &=& E\left\{ Z p_{1t}(X) + (1-Z)\frac{p_{1t}(X)}{\varepsilon_{1t}(X)} \right\}  \label{eq::survival-reg1} \\
    &=& E\left\{ w_{1t}(X) \frac{Z1(Y>t)}{e(X)} \right\}  \label{eq::survival-ipw1} \\
    &=& E\left\{ w_{1t}(X) \frac{Z1(Y>t)}{e(X)} - \frac{Z-e(X)}{e(X)}\frac{p_{1t}(X)}{\varepsilon_{1t}(X)}  \right\}, \label{eq::survival-dr1}
    \end{eqnarray}
    and
    \begin{eqnarray}
    S_0(t) &=& E\left\{ Z p_{0t}(X)\varepsilon_{0t}(X) + (1-Z)p_{0t}(X) \right\}  \label{eq::survival-reg0} \\
    &=& E\left\{ w_{0t}(X) \frac{(1-Z)1(Y>t)}{1-e(X)} \right\} \label{eq::survival-ipw0} \\
    &=& E\left\{ w_{0t}(X) \frac{(1-Z)1(Y>t)}{1-e(X)} - \frac{e(X)-Z}{1-e(X)}p_{0t}(X)\varepsilon_{0t}(X) \right\}\label{eq::survival-dr0}
\end{eqnarray}
where   $w_{1t}(X)=e(X)+\{1-e(X)\}/\varepsilon_{1t}(X)$ and $w_{0t}(X)=e(X)\varepsilon_{0t}(X)+1-e(X)$ are the weights.

Censoring is a major challenge for analyzing survival outcomes. To construct the outcome regression estimator, we must use models and estimators tailored to survival outcomes and censoring (e.g., \citealt{cox1972regression}). With estimated $p_{zt}(X)$, \eqref{eq::survival-reg1} and \eqref{eq::survival-reg0} motivate the outcome regression estimator 
\begin{eqnarray*}
    \hat{\tau}^{\textup{reg}}(t) &=& \left\{ n^{-1}\sumn Z_i \hat{p}_{1t}(X_i) + n^{-1} \sumn (1-Z_i)\hat{p}_{1t}(X_i)/\varepsilon_{1t}(X_i)\right\} \\
    && - \left\{ n^{-1} \sumn Z_i \hat{p}_{0t}(X_i)\varepsilon_{0t}(X_i) + n^{-1} \sumn (1-Z_i)\hat{p}_{0t}(X_i) \right\}.
\end{eqnarray*}
To construct the inverse propensity score weighting estimator, we must use an inverse probability weighting adjusted Kaplan--Meier estimator  \citep{xie2005adjusted}. Let  $\Tilde{Y}_i = \min (Y_i, C_i)$ denote the minimum of the outcome $Y_i$ and the censoring time $C_i$, and $\Delta_i = 1(Y_i \leq C_i)$ denote the censoring indicator. Let $t_j$'s denote the times when at least one event happens. For each group with treatment $z\in\{0,1\}$, define the inverse probability weighted number of events that happened at time $t_j$ and the inverse probability weighted number of individuals at risk at time $t_j$ as
\begin{eqnarray*}
    d_{j1} = \sum_{i:\Tilde{Y}_i = t_j} \frac{\hat{w}_{1t}(X_i) Z_i \Delta_i}{\hat{e}(X_i)}, &\ & n_{j1} = \sum_{i:\Tilde{Y}_i \geq t_j} \frac{\hat{w}_{1t}(X_i) Z_i}{\hat{e}(X_i)}, \\
    d_{j0} = \sum_{i:\Tilde{Y}_i = t_j} \frac{\hat{w}_{0t}(X_i) (1-Z_i) \Delta_i}{1-\hat{e}(X_i)}, &\ & n_{j0} = \sum_{i:\Tilde{Y}_i \geq t_j} \frac{\hat{w}_{0t}(X_i) (1-Z_i)}{1-\hat{e}(X_i)} .
\end{eqnarray*}
Then \eqref{eq::survival-ipw1} and \eqref{eq::survival-ipw0} motivate the weighted Kaplan--Meier estimator 
$$
\hat{\tau}^{\textup{ht}}(t) = \prod_{t_j \leq t} \left( 1 - \frac{d_{j1}}{n_{j1}} \right) - \prod_{t_j \leq t} \left( 1 - \frac{d_{j0}}{n_{j0}} \right).
$$
With the inverse probability weighting estimator, \eqref{eq::survival-dr1} and \eqref{eq::survival-dr0} motivate the following estimator that combines the estimated propensity score and the survival probability models
$$
\hat{\tau}^{\textup{dr}}(t) = \hat{\tau}^{\textup{ht}}(t) - n^{-1}\sumn \frac{\check{Z}_i \hat{p}_{1t}(X_i)}{\hat{e}(X_i)\varepsilon_{1t}(X_i)} - n^{-1}\sumn \frac{\check{Z}_i\hat{p}_{0t}(X_i)\varepsilon_{0t}(X_i)}{1-\hat{e}(X_i)}.
$$

\subsection{Extensions to observational studies with a multi-valued treatment}
\label{subsec::multi-valued}

The results also extend to observational studies with a multi-valued treatment, $Z\in \{1, \ldots, K\}$.  Each unit has $K$ potential outcomes $\{Y(1),\ldots,Y(K)\}$ corresponding to the $K$ treatment levels. We can define the causal parameters of interest as the comparisons of potential outcomes
$$
\tau_{c} = \sum_{k=1}^{K} c_k E\{Y(k)\},
$$
where $\sum_{k=1}^{K}c_k=0$.

Define the sensitivity parameters as below.

\begin{definition}
\label{def::multi-armed-sensitivity}
For any two treatment levels $k$ and $l$, define the sensitivity parameters as 
$$
\varepsilon_{k,l}(X) = \frac{E\{Y(k)\mid Z=k, X\}}{E\{Y(k) \mid Z=l, X\}}.
$$
\end{definition}

In Definition \ref{def::multi-armed-sensitivity}, we must have $\varepsilon_{k,k}(X)  = 1$ for all $k$s.
Under Definition \ref{def::multi-armed-sensitivity}, we have the following identification formulas for the mean of the potential outcomes
\begin{eqnarray*}
    E\{Y(k)\} &=& \sum_{l=1}^{K} E\left\{1(Z=l) \frac{\mu_k(X)}{\varepsilon_{k,l}(X)}\right\} \\
    &=& \sum_{l=1}^{K} E\left\{ \frac{e_l(X)}{\varepsilon_{k,l}(X)} \frac{1(Z=k)Y}{e_k(X)} \right\} \\
    &=& \sum_{l=1}^{K} E\left[ 1(Z=l)\frac{\mu_k(X)}{\varepsilon_{k,l}(X)} + \frac{e_l(X)}{\varepsilon_{k,l}(X)} \frac{1(Z=k)\{Y-\mu_k(X)\}}{e_k(X)} \right],
\end{eqnarray*}
where $e_k(X) = \pr(Z=k\mid X)$ is the generalized propensity score \citep{imbens2000role, imai2004causal} and $\mu_k(X)=E\{Y\mid Z=k, X\}$ is the conditional outcome mean. Motivated by the three identification formulas, we can construct the following estimators for $E\{Y(k)\}$,
\begin{eqnarray*}
    \hat{\mu}_k^{\textup{reg}} &=& n^{-1} \sumn \sum_{l=1}^{K} \frac{1(Z_i = l)\hat{\mu}_k(X_i)}{\varepsilon_{k,l}(X_i)} \\
    \hat{\mu}_k^{\textup{ht}} &=& n^{-1} \sumn \sum_{l=1}^{K} \frac{\hat{e}_l(X_i)1(Z_i=k)Y_i}{\varepsilon_{k,l}(X_i)\hat{e}_k(X_i)} \\
    \hat{\mu}_k^{\textup{dr}} &=& \hat{\mu}_k^{\textup{reg}} + n^{-1} \sumn \sum_{l=1}^{K} \frac{\hat{e}_l(X_i)1(Z_i=k)\{Y_i-\hat{\mu}_k(X_i)\}}{\varepsilon_{k,l}(X_i)\hat{e}_k(X_i)},
\end{eqnarray*}
and the following estimator for the causal parameter of interest 
$$
\hat{\tau}_{c}^{*} = \sum_{k=1}^{K} c_k \hat{\mu}_k^{*} \ (* = \textup{reg},\textup{ht},\textup{dr}).
$$

\subsection{Extensions to the controlled direct effect}

With an intermediate variable $S$ between the treatment $Z$ and the outcome $Y$, we use the notation $Y(z, s)$ to denote the potential outcome under treatment assignment $Z=z$ and intermediate value $S=s$. When $Z$ is binary, for a fixed level $s$, we define the controlled direct effect as $$\textsc{CDE}(s) = E\{Y(1,s)-Y(0,s)\},$$ which measures the average treatment effect of $Z$ when the level of the intermediate variable is fixed at $s$, thus measures the direct effect of treatment on the outcome controlling the mediator value at $s$. As an extension, if $S$ is discrete, we can view the problem of identifying and estimating the controlled direct effect as an observation study with multiple treatment levels (\citealp[]{vanderweele2015explanation}, \citealp[Chapter 28 in][]{ding2023first}). Therefore, we can apply the above framework in Section~\ref{subsec::multi-valued} to conduct sensitivity analysis for the controlled direct effect. We omit the details. 

\subsection{Formulas for the Hajek-type estimators}\label{subsec::hajek}
Under our sensitivity analysis framework, Theorem \ref{thm::ipw} motivates the following Hajek estimator of average causal effect
$$
\hat\tau^{\textup{haj}} =   \sumn \hat{w}_1(X_i)\frac{Z_i Y_i}{\hat{e}(X_i)} \Big/  \sumn \frac{Z_i}{ \hat{e}(X_i)  } - \sumn \hat{w}_0(X_i)\frac{(1-Z_i) Y_i}{1-\hat{e}(X_i)} \Big/  \sumn \frac{1-Z_i}{ 1- \hat{e}(X_i)  }.
$$
The denominators in the Hajek estimator take the same form as the canonical ones, which do not depend on $\varepsilon_1(X_i)$ or $\varepsilon_0(X_i)$. 
We also have the following alternative form of the doubly robust estimator based on the Hajek estimator
$$
\hat{\tau}^{\textup{dr2}} = \hat{\tau}^{\textup{proj}} + \sumn \hat{w}_1(X_i)\frac{Z_i \check{Y}_i}{\hat{e}(X_i)} \Big/  \sumn \frac{Z_i}{ \hat{e}(X_i)  } - \sumn \hat{w}_0(X_i)\frac{(1-Z_i) \check{Y}_i}{1-\hat{e}(X_i)} \Big/  \sumn \frac{1-Z_i}{ 1- \hat{e}(X_i)  }.
$$
In the special case with unconfoundedness assumption, $\varepsilon_1(X_i) = \varepsilon_0(X_i)=1$, they reduce to the standard Hajek estimator and doubly robust estimator for $\tau$. 

Similarly, Theorem \ref{thm::att-reg-ipw} motivates the following Hajek estimator of the average treatment effect on the treated units, $\hat{\tau}_{\textsc{t}}^\textup{haj} = \hat{\mu}_{\textsc{t}1}  - \hat{\mu}_{\textsc{t}0}^{\textup{haj}}$ where
$$
    \hat{\mu}_{\textsc{t}0}^\textup{haj} = \sumn   \varepsilon_0(X_i) \hat  o(X_i) (1-Z_i) Y_i  \Big/  \sumn   \hat  o(X_i) (1-Z_i).
$$
The doubly robust estimator is $\hat{\tau}_{\textsc{t}}^\textup{dr2} = \hat{\mu}_{\textsc{t}1}  - \hat{\mu}_{\textsc{t}0}^{\textup{dr2}}$ with
$$
    \hat{\mu}_{\textsc{t}0}^\textup{dr2} = \hat{\mu}_{\textsc{t}0}^{\text{reg}} + \sumn \varepsilon_0(X_i) \hat  o(X_i) (1-Z_i) \check{Y}_i  \Big/  \sumn   \hat  o(X_i) (1-Z_i).
$$
Again, in the special case when $\varepsilon_0(X_i)=1$, they reduce to the standard Hajek estimator and doubly robust estimator for $\tau_{\textsc{t}}$.

\subsection{Formulas for the bias-corrected matching estimator}\label{subsec::matching}

We also provide the formulas for the bias-corrected matching estimator under our sensitivity analysis framework. Matching estimators impute the missing counterfactual $Y_i(1-z)$ for an individual $i$ in treatment group $Z_i=z$ by finding $M$ nearest neighbors of $i$ in treatment group $Z_i=1-z$ and use the average observed outcomes of the $M$ nearest neighbors as the imputed value of the counterfactual $Y_i(1-z)$ for $z\in\{0,1\}$. The matching-based estimator of $\tau$ is then constructed by taking the average of the imputed individual treatment effect \citep{abadie2006large}. This estimator is generally not consistent when the nearest neighbor matching is based on multi-dimensional observed covariates. \cite{abadie2011bias} proposed a bias-corrected version of the matching estimator by estimating the conditional outcome models $\mu_z(X)$ and combining it with the original matching estimator. As shown in Lemma 5.1 in \cite{lin2021estimation}, under Assumption \ref{assume::unconfoundedness}, viewing matching as a nonparametric method of estimating the propensity score, we can rewrite the bias-corrected matching estimator in \cite{abadie2011bias} as
\begin{eqnarray*}
    \hat{\tau}^{\textup{bc}}_{M} &=& \hat{\tau}^{\textup{reg}} + n^{-1}\sumn\left[ \left\{1+\frac{K_M^{1}(X_i)}{M}\right\}Z_i\check{Y}_i - \left\{1+\frac{K_M^{0}(X_i)}{M}\right\}(1-Z_i)\check{Y}_i\right],
\end{eqnarray*}
where $M$ is the fixed number of matches for each observation and $K_M^{z}(X_i)$ is the number of matched times of unit $i$ in treatment group $z$, $z\in\{0,1\}$. 

Under our sensitivity analysis framework, we can rewrite the bias-corrected matching estimator as
\begin{eqnarray*}
    \hat{\tau}^{\textup{bc}}_{M} &=& \hat{\tau}^{\textup{pred}} + n^{-1}\sumn \left\{ \frac{K_M^1(X_i)}{M} \frac{1}{\varepsilon_1(X_i)} Z_i \check{Y}_i - \frac{K_M^0(X_i)}{M} \varepsilon_0(X_i)(1-Z_i)\check{Y}_i \right\}.
\end{eqnarray*}
\cite{lin2021estimation} shows that under some regularity conditions, $n_zK_M^z(X_i)/(n_{1-z}M)$ consistently estimates the density ratio $f_{X\mid Z=1-z}(X_i)/f_{X\mid Z=z}(X_i)$ for $z\in\{0,1\}$, where $n_z$ is the total number of observations in treatment group $Z=z$. Moreover, $n_1/n_0$ is a consistent estimator of the ratio of the marginal probabilities $\pr(Z=1)/\pr(Z=0)$, thus we essentially use $1+K_M^1(X_i)/M$ and $1+K_M^0(X_i)/M$ to estimate $1/e(X_i)$ and $1/\{1-e(X_i)\}$, respectively.

\subsection{Semiparametric efficiency bounds for $\tau$ and $\tau_\T$}\label{subsec::semi_efficiency_bound}
In this subsection, we provide the semiparametric efficiency bound for $\tau$ and $\tau_\T$ under our sensitivity analysis framework in the following proposition.
\begin{proposition}
    Under Definition~\ref{def::sensitivity-analysis}, the semiparametric efficiency bounds for $\tau$ and $\tau_\T$ are
    \begin{eqnarray*}
        \var(\phi) &=& E\left\{\frac{w_1^2(X)}{e(X)}\var(Y\mid X,Z=1)+\frac{w_0^2(X)}{1-e(X)}\var(Y\mid X,Z=0)\right\} \\
        &&+ E\left[e(X)\left\{\mu_1(X)-\mu_0(X)\varepsilon_0(X)\right\}^2 + \left\{1-e(X)\right\}\left\{\frac{\mu_1(X)}{\varepsilon_1(X)}-\mu_0(X)\right\}^2 - \tau^2\right],
    \end{eqnarray*}
    and
    \begin{eqnarray*}
        \var(\phi_\T) &=& E\left[\frac{e(X)\var(Y\mid X,Z=1)}{e^2}+\frac{e^2(X)\varepsilon_0^2(X)\var(Y\mid X,Z=0)}{e^2\left\{1-e(X)\right\}}\right] \\
        &&+ E\left[\frac{e(X)\left\{\mu_1(X)-\varepsilon_0(X)\mu_0(X)-\tau_\T\right\}^2}{e^2}\right],
    \end{eqnarray*}
    respectively.
\label{prop:semiparametric_bound}
\end{proposition}
Proposition~\ref{prop:semiparametric_bound} shows that the semiparametric efficiency bound for $\tau$ is not monotonic in $\varepsilon_1(X)$ and $\varepsilon_0(X)$ even when they are constant, and similarly that for $\tau_\T$ is not monotonic in $\varepsilon_0(X)$ either. This echoes our discussion in Section~\ref{subsec::application}.

\subsection{Connection to the treatment selection model}\label{subsec::treatment_selection}
In this subsection, we show that our proposed sensitivity parameters are also related to the treatment selection model, although their definitions are based on the outcome models. 

Take the treatment selection model $\pr\{Z=1\mid X, Y(1)\}$ for example. Under Definition~\ref{def::sensitivity-analysis}, when $\varepsilon_1(X) \neq 1$, $\pr\{Z=1\mid X, Y(1)\}$ is generally not equal to $e(X)$ and does depend on $Y(1)$. We next derive the relationship between $\pr\{Z=1\mid X, Y(1)\}$ and $e(X)$ to provide the connection between our sensitivity parameters and the selection model. 

We start with the special case when $Y$ is binary. In this case, we have
\begin{eqnarray}
    e(X) &=& e(X,1)w_1(X)\mu_1(X) + e(X,0)\left\{1-w_1(X)\mu_1(X)\right\},\label{eqn::selection_binary_y}
\end{eqnarray}
where $e(X,y) = E\{Z\mid X, Y(1)=y\}$ is the propensity score under selection to treatment for $y=0,1$. The $w_1(X)=e(X) + \{1-e(X)\}/ \varepsilon_1(X)$ term on the right-hand side depends on $\varepsilon_1(X)$, thus~\eqref{eqn::selection_binary_y} shows the connection between the sensitivity parameter and the selection model. Previous sensitivity analysis framework bounds the ratio $$\frac{\pr\{Z=1\mid X,Y(1)\} / [1-\pr\{Z=1\mid X,Y(1)\}]}{e(X)/\{1-e(X)\}}$$ to estimate $E\{Y(1)\}$, which works better for the IPW estimators \citep{zhao2019sensitivity, dorn2022sharp}.

For a general $Y$, we first introduce a set of $\varepsilon_1(X,y)$ to denote the ratio between the observed and counterfactual conditional densities, 
\begin{eqnarray*}
    \frac{p_{Y(1)\mid X,Z=1}(y\mid X, Z=1)}{p_{Y(1)\mid X,Z=0}(y\mid X, Z=0)} &=& \varepsilon_1(X,y),
\end{eqnarray*}
where $p_{Y(1)\mid X,Z=z}(y\mid X, Z=z)$ denotes the conditional density of $Y(1)$ in treatment group $Z=z$ for $z=0,1$ with $\int y\left\{1-\varepsilon_1(X)/\varepsilon_1(X,y)\right\}p_{Y(1)\mid X,Z=1}(y\mid X, Z=1)\textup{d}y = 0$. We have the following equality
\begin{eqnarray*}
    e(X) &=& \int e(X,y)\left\{e(X) + \frac{1-e(X)}{\varepsilon_1(X,y)}\right\}p_{Y(1)\mid X,Z=1}(y\mid X, Z=1)\textup{d}y.
\end{eqnarray*}
This provides the relation of treatment selection model $\pr\{Z=1\mid X, Y(1)\}$ with our sensitivity parameter.

\section{R package and implementation}\label{sec::R-package}
We provide an R package \texttt{saci} to conduct the proposed sensitivity analysis. The package can be installed from Github using the command: \texttt{devtools::install\_github("sizhu-lu/saci")}.
We provide the following functions in this package:
\begin{lstlisting}
sa_ate(X, Z, Y, eps1_list, eps0_list,
       pscore_family="binomial", outcome_family="gaussian",
       n_boot=500, truncpscore=c(0,1))
\end{lstlisting}

\texttt{sa\_ate} is used to conduct sensitivity analysis of the average treatment effect. The inputs include the observed data, the user-specified lists of $\varepsilon_1$ and $\varepsilon_0$, the specification of the ``family'' option in the generalized linear models of propensity score model and outcome model, the total number of bootstrap samples to compute the estimated standard errors, and the truncation cutoffs of the estimated propensity score. This function outputs the point estimator, estimated standard error, p-value, and the 95\% confidence interval for each pair of sensitivity parameters $(\varepsilon_1, \varepsilon_0)$ in the user-specified lists and for each method (proj, ht, hajek, and dr). We use the nonparametric bootstrap for standard error estimation.

\begin{lstlisting}
sa_att(X, Z, Y, eps0_list,
       pscore_family="binomial", outcome_family="gaussian",
       n_boot=500, truncpscore=c(0,1))
\end{lstlisting}

\texttt{sa\_att} is used to conduct sensitivity analysis of the average treatment effect on the treated. The inputs and outputs are similar to \texttt{sa\_ate}, except that the estimation procedure only depends on $\varepsilon_0$.

\begin{lstlisting}
sa_rr(X, Z, Y, eps1_list, eps0_list,
       pscore_family="binomial", outcome_family="binomial",
       n_boot=500, truncpscore=c(0,1), log="FALSE")
\end{lstlisting}

For binary potential outcomes, the function \texttt{sa\_rr} is used to conduct sensitivity analysis of the causal risk ratio or log causal risk ratio. The inputs include the observed data, the user-specified list of $\varepsilon_1$ and $\varepsilon_0$, the specification of the ``family'' option in the generalized linear models of propensity score model and outcome model, the total number of bootstrap samples to compute the estimated standard errors, the truncation cutoffs of the estimated propensity score, and option to take the logarithm of the causal risk ratio or not. This function outputs the point estimator, estimated standard error, p-value, and the 95\% confidence interval for each pair of sensitivity parameters $(\varepsilon_1, \varepsilon_0)$ in the user-specified lists and for each method (proj, ht, hajek, and dr). Again, we use the nonparametric bootstrap for standard error estimation.

\begin{lstlisting}
sa_or(X, Z, Y, eps1_list, eps0_list,
       pscore_family="binomial", outcome_family="binomial",
       n_boot=500, truncpscore=c(0,1), log="FALSE")
\end{lstlisting}

For binary potential outcomes, the function \texttt{sa\_or} is used to conduct sensitivity analysis of the causal odds ratio or log causal odds ratio. The inputs and outputs are similar to \texttt{sa\_rr}.

We also provide a function \texttt{plot\_contour} to generate a contour plot of the estimated average treatment effects with different values of sensitivity parameters. The usage is
\begin{lstlisting}
plot_contour(ate_result, caption="", estimator="dr", value="est")
\end{lstlisting}
where \texttt{ate\_result} is the output of the \texttt{sa\_ate} function and the default value to plot is the doubly robust point estimator. Figure \ref{fig::contour_plot} shows a sample plot for the observational study in Section~\ref{subsec::application}. Users are also free to change the estimation method to other estimators and the plotted values to lower or upper confidence bounds by specifying, e.g., \texttt{estimator="ht"} and \texttt{value="ci\_lb"}.

\section{Proofs}\label{sec::proofs}

\subsection{Proof of Theorem \ref{thm::outcome-reg-sensitivity}}
We have 
\begin{eqnarray*}
E\{ Y(1)\mid Z=0 \}  
&=& E\left[ E\{ Y(1)\mid Z=0, X \}  \mid Z=0 \right]  \\
&=& E\left[ E\{ Y(1)\mid Z=1, X \} /\varepsilon_1(X)   \mid Z=0\right]  \\
&=&  E\left\{\mu_1(X)   / \varepsilon_1(X)  \mid Z=0\right\}
\end{eqnarray*}
and
\begin{eqnarray*}
E\{ Y(0)\mid Z=1 \}  
&=& E\left[ E\{ Y(0)\mid Z=1, X \}  \mid Z=1 \right]  \\
&=& E\left[ E\{ Y(0)\mid Z=0, X \} \varepsilon_0(X)   \mid Z=1 \right]  \\
&=&  E\left\{\mu_0(X)    \varepsilon_0(X)  \mid Z=1\right\}.
\end{eqnarray*}
Therefore, we can write $\tau$ as the difference between 
\begin{eqnarray*}
E\{ Y(1) \} &=&   E(Y\mid Z=1) \pr(Z=1) + E\left\{\mu_1(X)   / \varepsilon_1(X)  \mid Z=0\right\} \pr(Z=0)   ,\\
 E\{Y(0)\} &=&   E\left\{\mu_0(X)    \varepsilon_0(X)  \mid Z=1\right\}\pr(Z=1) + E(Y\mid Z=0)  \pr(Z=0)  ,
\end{eqnarray*}
which can be simplified as the difference between 
\begin{eqnarray*}
E\{ Y(1) \} =   E\{ ZY + (1-Z) \mu_1(X)   / \varepsilon_1(X)\}  ,\quad
 E\{Y(0)\} =    E\{ Z \mu_0(X)    \varepsilon_0(X) + (1-Z)Y\}. 
\end{eqnarray*}
This gives \eqref{eq::projective}. 
The identification formula \eqref{eq::predictive} follows from  
$$
E(ZY) = E(Y\mid Z=1)\pr(Z=1) = E\{  \mu_1(X)\mid Z=1 \} \pr(Z=1)
$$
and
$$
E\{(1-Z)Y\} = E(Y\mid Z=0)\pr(Z=0)=E\{  \mu_0(X)\mid Z=0 \} \pr(Z=0).
$$

\subsection{Proof of Theorem \ref{thm::ipw}}
We only prove the result under treatment because  the proof for the result under control is similar. On the one hand, we can use \eqref{eq::predictive} to obtain
\begin{eqnarray*} 
E\{ Y(1) \}  &=&    E\{ ZY + (1-Z) \mu_1(X)   / \varepsilon_1(X)\}  \\
&=& E\left[  e(X) \mu_1(X) + \{ 1-e(X)\} \mu_1(X)   / \varepsilon_1(X) \right]  \\
&=& E\{ w_1(X)  \mu_1(X) \} .
\end{eqnarray*}
On the other hand, we can condition on $X$ to obtain 
\begin{eqnarray*}
E\left\{ w_1(X) \frac{Z}{e(X)}  Y  \right\}  &=& E\left\{ w_1(X) \frac{  E(Z\mid X) }{e(X)}  E(Y\mid Z=1, X)  \right\} \\
&=& E \{ w_1(X)  \mu_1(X)   \} .
\end{eqnarray*}
The conclusion for $Y(1)$ follows.

\subsection{Proof of Theorem \ref{thm::eif}}\label{subsec::proof_eif}
We first present a short but informal proof.
Based on the form $E\{ Y(1) \}  = E\{ w_1(X)  \mu_1(X) \}$ as a byproduct in the proof of Theorem \ref{thm::ipw}, we can derive the efficient influence function for $E\{ Y(1)\}$ by extending the result of \cite{hahn1998role}. The key difference is that the weight depends on the propensity score, so we must include an additional term due to the derivative of $w_1(X) $ with respect to $e(X)$. This additional term comes from the path-wise derivative $\dot e_\theta(X)|_{\theta=0} = E [  \{ Z-e(X)\} s(Z\mid X) \mid X ]$, where $\theta$ is the parameter for the sub-model and $s(Z\mid X)$ is the score function for $E(Z\mid X)$. 
Including this additional term yields the following efficient influence function for $E\{ Y(1)\}$:
\begin{eqnarray*}
&& w_1(X) \left\{  \frac{Z}{e(X)}  Y  - \frac{Z-e(X)}{e(X)} \mu_1(X)       \right\}  + \frac{\varepsilon_1(X) - 1}{\varepsilon_1(X)} \{ Z- e(X)\} \mu_1(X) \\
&=& w_1(X)  \frac{Z}{e(X)}  Y  -  \left\{  \frac{w_1(X)}{ e(X) } -  \frac{\varepsilon_1(X) - 1}{\varepsilon_1(X)}   \right\} \{Z-e(X)\}\mu_1(X)  , 
\end{eqnarray*}
which reduces to $\phi_1 $ by simple algebra. The efficient influence function for $E\{ Y(0)\}$ follows from a similar calculation.

We then provide a longer but formal proof.
We follow the semiparametric theory in \cite{bickel1993efficient} and \cite{hahn1998role} to derive the efficient influence function for $E\left\{ Y(1)\right\} $. We denote the vector of all observed variables $V=(X,Z,Y)$ and factorize the likelihood as 
\[
P(V)=P(X)P(Z\mid X)P(Y\mid Z,X).
\]
To derive the efficient influence function for $E\left\{ Y(1)\right\}$, we consider a one-dimensional parametric submodel $P_{\theta}\left(V\right)$ which contains the true model $P(V)$ at $\theta=0$. We use $\theta$ in the subscript to denote the quantity with respect to the submodel, e.g., $e_{\theta}\left(s_{1},s_{0}\right)$ is the value of $e\left(s_{1},s_{0}\right)$ with respect to the submodel. We use dot to denote the partial derivative with respect to $\theta$, e.g., $\dot{e}_{\theta}\left(s_{1},s_{0}\right) = \partial e_{\theta}\left(s_{1},s_{0}\right) / \partial \theta$, and use $s_{\theta}(\cdot)$ to denote the score function with respect to the submodel. Decomposition of the score function under the submodel as 
\[
s_{\theta}(V)=s_{\theta}(X)+s_{\theta}(Z\mid X)+s_{\theta}(Y\mid Z,X)
\]
where $s_{\theta}(X)=\partial\log P_{\theta}(X)/\partial\theta$,
$s_{\theta}(Z\mid X)=\partial\log P_{\theta}(Z\mid X)/\partial\theta$,
and $s_{\theta}(Y\mid Z,X)=\partial\log P_{\theta}(Y\mid Z,X)/\partial\theta$
are the score functions corresponding to the three components of the
likelihood function. We write $\left.s_{\theta}(\cdot)\right|_{\theta=0}$ as $s(\cdot)$, which is the score function evaluated at the true model $\theta=0$ under the one-dimensional submodel.   Let $\phi_{1}\left(V\right)$ denote the efficient influence function for $E\left\{ Y(1)\right\} $.
It must satisfy the equation 
$$
\left.\frac{\partial E_{\theta}\left\{ Y(1)\right\} }{\partial\theta}\right|_{\theta=0}=E\left\{ \phi_{1}(V)s(V)\right\} .
$$
To find such $\phi_{1}(V)$, we need to calculate the derivative of
\allowdisplaybreaks
\begin{eqnarray*}
    E_{\theta}\left\{ Y\left(1\right)\right\} 
    &=&E_{\theta}\left\{ w_{1}\left(X\right)\mu_{1}\left(X\right)\right\} \\
    &=&\int\frac{e_{\theta}\left(x\right)\varepsilon_{1}\left(x\right)+1-e_{\theta}\left(x\right)}{\varepsilon_{1}\left(x\right)}\mu_{1,\theta}\left(x\right)dP_{\theta}\left(x\right).
\end{eqnarray*}
Therefore, the Gateaux derivative of our parameter of interest under the parametric submodel with respect to $\theta$ is 
\begin{eqnarray*}
    \left.\frac{\partial E_{\theta}\left\{ Y(1)\right\} }{\partial\theta}\right|_{\theta=0}&=&\int\frac{\left.\dot{e}_{\theta}\left(x\right)\right|_{\theta=0}\varepsilon_{1}\left(x\right)-\left.\dot{e}_{\theta}\left(x\right)\right|_{\theta=0}}{\varepsilon_{1}\left(x\right)}\mu_{1}\left(x\right)dP\left(x\right)\\&&+\int w_{1}\left(x\right)\left.\dot{\mu}_{1,\theta}\left(x\right)\right|_{\theta=0}dP\left(x\right)\\&&+\int\left\{ w_{1}\left(x\right)\mu_{1}\left(x\right)-\mu_{1}\right\} s\left(x\right)dP\left(x\right)\\&=&E\left[ \left.\dot{e}_{\theta}\left(X\right)\right|_{\theta=0}\left\{1-\frac{1}{\varepsilon_{1}\left(X\right)}\right\}\mu_{1}\left(X\right)\right] \\&&+E\left\{ w_{1}\left(X\right)\left.\dot{\mu}_{1,\theta}\left(X\right)\right|_{\theta=0}\right\} \\&&+E\left[ \left\{w_{1}\left(X\right)\mu_{1}\left(X\right)-\mu_{1}\right\}s\left(X\right)\right] .
\end{eqnarray*}
The calculation of two building blocks $\left.\dot{e}_{\theta}\left(X\right)\right|_{\theta=0}$ and $\left.\dot{\mu}_{1,\theta}\left(X\right)\right|_{\theta=0}$ are standard \citep{hahn1998role}:
\begin{eqnarray*}
    \left.\dot{e}_{\theta}\left(X\right)\right|_{\theta=0}&=&\left.\frac{\partial E_{\theta}\left(Z\mid X\right)}{\partial\theta}\right|_{\theta=0}=E\left[ \left\{Z-e\left(X\right)\right\}s\left(Z\mid X\right)\mid X\right], \\
    \left.\dot{\mu}_{1,\theta}\left(X\right)\right|_{\theta=0}&=&\left.\frac{\partial E_{\theta}\left(Y\mid Z=1,X\right)}{\partial\theta}\right|_{\theta=0}\\
    &=&E\left[ \left\{Y-\mu_{1}\left(X\right)\right\}s\left(Y\mid Z=1,X\right)\mid Z=1,X\right] \\
    &=&E\left[ \frac{Z\left\{Y-\mu_{1}\left(X\right)\right\}}{e\left(X\right)}s\left(Y\mid Z,X\right)\mid X\right].
\end{eqnarray*}
Plugging back to the previous equation, we have
\begin{eqnarray*}
    &&\left.\frac{\partial E_{\theta}\left\{ Y(1)\right\} }{\partial\theta}\right|_{\theta=0}\\
    &=&E\left[ \left\{ Z-e\left(X\right)\right\}\left\{1-\frac{1}{\varepsilon_{1}\left(X\right)}\right\}\mu_{1}\left(X\right)s\left(Z\mid X\right)\right] \\
    &&+E\left[ w_{1}\left(X\right)\frac{Z\left\{Y-\mu_{1}\left(X\right)\right\}}{e\left(X\right)}s\left(Y\mid Z,X\right)\right] \\
    &&+E\left[\left\{ w_{1}\left(X\right)\mu_{1}\left(X\right)-\mu_{1}\right\}s\left(X\right)\right] \\
    &=&E\left(\left[ \left\{ Z-e\left(X\right)\right\}\left\{1-\frac{1}{\varepsilon_{1}\left(X\right)}\right\}\mu_{1}\left(X\right) +w_{1}\left(X\right)\frac{Z\left\{Y-\mu_{1}\left(X\right)\right\}}{e\left(X\right)}+w_{1}\left(X\right)\mu_{1}\left(X\right)-\mu_{1}\right]s\left(V\right)\right) \\
    &=&E\left( \left[w_{1}\left(X\right)\frac{ZY}{e\left(X\right)}-\frac{\left\{Z-e\left(X\right)\right\}\mu_{1}\left(X\right)}{e\left(X\right)\varepsilon_{1}\left(X\right)}-\mu_{1}\right]s\left(V\right)\right) ,
\end{eqnarray*}
where the second equality is by the properties of score functions and the last equality is by rearranging terms. Thus,
\begin{equation*}
    \phi_1\left(V\right) = w_{1}\left(X\right)\frac{ZY}{e\left(X\right)}-\frac{\left\{Z-e\left(X\right)\right\}\mu_{1}\left(X\right)}{e\left(X\right)\varepsilon_{1}\left(X\right)}-\mu_1.
\end{equation*}
We can prove 
\begin{equation*}
    \phi_0\left(V\right) = w_0(X) \frac{(1-Z)Y}{1-e(X)} - \frac{ \{e(X) - Z\} \mu_0(X)\varepsilon_0(X) }{ 1-e(X) } - \mu_0
\end{equation*}
using a similar calculation.

\subsection{Proof of Theorem \ref{thm::dr}}
If the fitted propensity score model converges to $e^*(X)$ and the fitted outcome models converge to $\mu_1^*(X)$ and $\mu_0^*(X)$, the doubly robust estimator has limit $\tau^{\textup{dr}}  = \mu_1^{\textup{dr}} - \mu_0^{\textup{dr}}$, where
\begin{eqnarray*}
\mu_1^{\textup{dr}}  = E\left[   \left\{   e^*(X) + \frac{1 - e^*(X) }{ \varepsilon_1(X)  }   \right\}   \frac{ZY}{ e^*(X) } \right] 
- E\left\{\frac{Z -e^*(X)}{e^*(X)}    \frac{ \mu_1^*(X)   }{  \varepsilon_1(X)  }   \right\} ,
\end{eqnarray*}
and the formula for $\mu_0^{\textup{dr}}$ is omitted due to similarity. We will only show that $\mu_1^{\textup{dr}} = E\{ Y(1) \}$ if either the propensity score or the outcome model is correctly specified. The proof for the control potential outcome is analogous.

If $e^*(X) = e(X)$, the conclusion is obvious since the inverse probability weighting estimator is consistent and the additional term has mean zero. If $\mu_1^*(X) = \mu_1(X)$, then by definition and the law of iterated expectations, $\mu_1^{\textup{dr}}  - E\{ Y(1) \} $ equals the expectation of 
$$
  \frac{\mu_1(X)}{\varepsilon_1(X)e^*(X)}\Delta(X)
$$
where
\begin{eqnarray*}
\Delta(X) &=& 
\{ e^*(X) \varepsilon_1(X) +1 - e^*(X) \} e(X) - \{ e(X) -  e^*(X)\}
- \{ e(X) \varepsilon_1(X) +1 - e(X) \} e^*(X)  \\
&=& e^*(X) \varepsilon_1(X) e(X)  +e(X)  - e^*(X)   e(X) - e(X) + e^*(X) \\
& & - e(X) \varepsilon_1(X) e^*(X) - e^*(X)  + e(X) e^*(X)   \\
&=& 0. 
\end{eqnarray*}
Therefore, if $\mu_1^*(X) = \mu_1(X)$, then $\mu_1^{\textup{dr}}  =  E\{ Y(1) \} $. 

\subsection{Proof of Proposition \ref{prop::bound_constant_epsilons}}
Recall that $\varepsilon_{z,\textsc{l}}$ and $\varepsilon_{z,\textsc{u}}$ are the lower and upper bound of $\varepsilon_z(X)$, i.e., $\varepsilon_z(X)\in[\varepsilon_{z,\textsc{l}}, \varepsilon_{z,\textsc{u}}]$ for $z=0,1$.
First, when $Y\geq 0$, we have $\mu_z(X)=E(Y\mid X,Z=z)\geq 0$ for $z=0,1$. Therefore, $(1-Z)\mu_1(X)\geq 0$ and $Z\mu_0(X)\geq 0$. By the identification formula in Theorem~\ref{thm::outcome-reg-sensitivity}, we have
\begin{eqnarray*}
    \tau & \geq & E\left\{Z\mu_1(X) + \frac{(1-Z)\mu_1(X)}{\varepsilon_{1,\textsc{u}}} - Z\mu_0(X)\varepsilon_{0,\textsc{u}} - (1-Z)\mu_0(X)\right\}.
\end{eqnarray*}

Second, when $Y\geq 0$, we have $\{1-e(X)\}ZY/e(X)\geq 0$ and $e(X)(1-Z)Y/\{1-e(X)\} \geq 0$. Therefore, by the identification formula in Theorem~\ref{thm::ipw}, we have
\begin{eqnarray*}
    \tau & \geq & E\left[\left\{e(X) + \frac{1-e(X)}{\varepsilon_{1,\textsc{u}}}\right\}\frac{ZY}{e(X)} - \left\{\varepsilon_{0,\textsc{u}}+1-e(X)\right\}\frac{(1-Z)Y}{1-e(X)}\right].
\end{eqnarray*}

We prove the third formula by showing the third lower bound is equal to the previous two bounds if either the propensity score or the outcome model is correctly specified. If the propensity score model is correct, we have
\begin{eqnarray*}
    E\left[\frac{\left\{Z-e(X)\right\}\mu_1(X)}{e(X)}\right] = E\left[\frac{\left\{Z-e(X)\right\}\mu_0(X)}{1-e(X)}\right] = 0.
\end{eqnarray*}
If the outcome model is correct, we have 
\begin{eqnarray*}
    && E\left[\left\{e(X) + \frac{1-e(X)}{\varepsilon_{1,\textsc{u}}}\right\}\frac{ZY}{e(X)}-\frac{\left\{Z-e(X)\right\}\mu_1(X)}{\varepsilon_{1,\textsc{u}}e(X)}\right] \\
    &=& E\left[ZY+\frac{\left\{1-e(X)\right\}ZY}{\varepsilon_{1,\textsc{u}}e(X)}-\frac{\left\{Z-e(X)\right\}\mu_1(X)}{\varepsilon_{1,\textsc{u}}e(X)}\right] \\
    &=& E\left[Z\mu_1(X) + \frac{(1-Z)\mu_1(X)}{\varepsilon_{1,\textsc{u}}}\right] \\
    &&+ E\left[Z\left\{Y-\mu_1(X)\right\} - \frac{(1-Z)\mu_1(X)e(X)- \left\{1-e(X)\right\}ZY + \left\{Z-e(X)\right\}\mu_1(X)}{\varepsilon_{1,\textsc{u}}e(X)}\right] \\
    &=& E\left[Z\mu_1(X) + \frac{(1-Z)\mu_1(X)}{\varepsilon_{1,\textsc{u}}}\right] + E\left[\left\{1 + \frac{1-e(X)}{\varepsilon_{1,\textsc{u}}e(X)}\right\}Z\left\{Y-\mu_1(X)\right\}\right] \\
    &=& E\left[Z\mu_1(X) + \frac{(1-Z)\mu_1(X)}{\varepsilon_{1,\textsc{u}}}\right],
\end{eqnarray*}
where the last equality follows from the fact that $\mu_1(X)=E(Y\mid Z=1,X)$. Similarly, we have
\begin{eqnarray*}
    && E\left[\left\{\varepsilon_{0,\textsc{u}}e(X)+1-e(X)\right\}\frac{(1-Z)Y}{1-e(X)}-\frac{\left\{e(X)=Z\right\}\mu_0(X)\varepsilon_{0,\textsc{u}}}{1-e(X)}\right] \\
    &=& E\left\{Z\mu_0(X)\varepsilon_{0,\textsc{u}} + (1-Z)\mu_0(X)\right\}.
\end{eqnarray*}
Therefore, the third formula of the lower bound is equal to the first two formulas if either the propensity score or the outcome model is correct.

The upper bound for $\tau$ can be derived similarly.


\subsection{Proof of Theorem \ref{thm::att-reg-ipw}}
On the one hand, we have
 \begin{eqnarray*}
E\{  Y(0)  \mid Z=1  \}  &=& E\big[   E\{ Y(0) \mid Z=1, X \}  \mid Z=1  \Big ] \\
&=& E\left\{   \mu_0(X)  \varepsilon_0(X)  \mid Z=1  \right\} \\
&=& E\left\{  Z \mu_0(X)  \varepsilon_0(X)    \right\} /e.
\end{eqnarray*}
Conditioning on $X$, we can simplify it to
$
E\{  Y(0)  \mid Z=1  \}  = E\left\{  e(X) \mu_0(X)  \varepsilon_0(X)    \right\} /e.
$

On the other hand, we condition on $X$ to obtain 
$$
E\left\{  e(X)\varepsilon_0(X) \frac{ 1-Z}{1-e(X)}  Y      \right\} = E\left\{  e(X)\varepsilon_0(X) E (Y\mid Z=0, X)      \right\} 
=  E\left\{  e(X)\varepsilon_0(X) \mu_0(X)   \right\} .
$$
Then the conclusion follows. 

\subsection{Proof of Theorem \ref{thm::eif-att}}

We first present a short but informal proof.  As a byproduct of the Proof of Theorem \ref{thm::att-reg-ipw}, we have $  E\{ Z Y(0)    \} = E\left\{    \varepsilon_0(X)  e(X)  \mu_0(X)   \right\}  $, which is a weighted average of $\mu_0(X)$. The result follows from Hahn with an additional term due to the derivative of the weight with respect to $e(X)$:
  \begin{eqnarray*}
 &&   \varepsilon_0(X)  e(X)   \left\{   \frac{1-Z}{1-e(X)} Y -  \frac{e(X) - Z}{1-e(X)}  \mu_0(X)  \right\}
  +  \varepsilon_0(X)  \{  Z-e(X) \}  \mu_0(X) \\
 &=&  \varepsilon_0(X)  e(X)    \frac{1-Z}{1-e(X)} Y - \varepsilon_0(X)  \{e(X) - Z\}  \mu_0(X)   \left\{  \frac{e(X)}{1-e(X)}   +1\right\} ,
 \end{eqnarray*}
 which reduces to the formula of $\phi_{\textsc{t}0} $ by simple algebra.

We then present a long but formal proof. 
We follow the same framework and notations as in the proof of Theorem \ref{thm::eif}. Again, we have $\mu_{\textsc{t}0}=E\left\{ ZY(0)\right\} /e = E\left\{    \varepsilon_0(X)  e(X)  \mu_0(X)   \right\}/e  $ as a byproduct of the Proof of Theorem \ref{thm::att-reg-ipw}. For a parametric submodel with parameter $\theta$, we have
\begin{eqnarray*}
\left.\frac{\partial}{\partial\theta}E_{\theta}\left\{ ZY(0)\right\} \right|_{\theta=0}&=&\left.\frac{\partial}{\partial\theta}\left\{ \int\varepsilon_{0}(x)e_{\theta}(x)\mu_{0,\theta}(x)p_{\theta}(x)dx\right\} \right|_{\theta=0}\\&=&E\left\{ \varepsilon_{0}(X)\left.\dot{e}_{\theta}(X)\right|_{\theta=0}\mu_{0}(X)\right\} \\&&+E\left\{ \varepsilon_{0}(X)e(X)\left.\dot{\mu}_{0,\theta}(X)\right|_{\theta=0}\right\} \\&&+E\left\{ \varepsilon_{0}(X)e(X)\mu_{0}(X)s(X)\right\} \\&=&E\left[ \varepsilon_{0}(X)\left\{Z-e\left(X\right)\right\}\mu_{0}(X)s\left(Z\mid X\right)\right] \\&&+E\left[ \varepsilon_{0}(X)e(X)\frac{\left(1-Z\right)\left\{Y-\mu_{0}\left(X\right)\right\}}{1-e\left(X\right)}s\left(Y\mid Z,X\right)\right] \\&&+E\left[\left\{ \varepsilon_{0}(X)e(X)\mu_{0}(X)-e\mu_{\textsc{t}0}\right\}s(X)\right] \\&=&E\left(\left[ Z\varepsilon_{0}(X)\mu_{0}(X)+\varepsilon_{0}(X)e(X)\frac{\left(1-Z\right)\left\{Y-\mu_{0}\left(X\right)\right\}}{1-e(X)}-e\mu_{\textsc{t}0}\right] s(V)\right),
\end{eqnarray*}
by the facts that
\begin{eqnarray*}
\left.\dot{e}_{\theta}\left(X\right)\right|_{\theta=0} &=& E\left\{ \left(Z-e\left(X\right)\right)s\left(Z\mid X\right)\mid X\right\}, \\
\left.\dot{\mu}_{0,\theta}(X)\right|_{\theta=0}&=&\left.\frac{\partial E_{\theta}\left(Y\mid Z=0,X\right)}{\partial\theta}\right|_{\theta=0}\\
&=&E\left\{ \left(Y-\mu_{0}\left(X\right)\right)s\left(Y\mid Z=0,X\right)\mid Z=0,X\right\} \\
&=&E\left[ \frac{\left(1-Z\right)\left\{Y-\mu_{0}\left(X\right)\right\}}{1-e\left(X\right)}s\left(Y\mid Z,X\right)\mid X\right],
\end{eqnarray*}
and the properties of the score functions. 

We also have that $\left.\dot{e}_{\theta}\right|_{\theta=0}=E\left\{ \left(Z-e\right)s\left(X\right)\right\} $. Therefore, the efficient influence function for $ \mu_{\textsc{t}0} $ is
\begin{eqnarray*}
    \phi_{\textsc{t}0}&=&e^{-1}\left[ Z\varepsilon_{0}(X)\mu_{0}(X)+\varepsilon_{0}(X)e(X)\frac{\left(1-Z\right)\left\{Y-\mu_{0}\left(X\right)\right\}}{1-e(X)}-e\mu_{\textsc{t}0}\right] - e^{-2}\left\{ e\mu_{\textsc{t}0}\left(Z-e\right)\right\} \\&=& e^{-1}\left[ Z\left\{\varepsilon_{0}(X)\mu_{0}(X)-\mu_{\textsc{t}0}\right\}+\varepsilon_{0}(X)e(X)\frac{\left(1-Z\right)\left\{Y-\mu_{0}\left(X\right)\right\}}{1-e(X)}\right].
\end{eqnarray*}
This concludes the proof.

\subsection{Proof of Theorem \ref{thm::dr-att}}
If the fitted propensity score model converges to $e^*(X)$ and the fitted outcome model under control converge to $\mu_0^*(X)$, the doubly robust estimator has limit $\tau_{\textsc{t}}^{\textup{dr}}  = E\{ Y(1) \mid Z=1\}- \mu_{\textsc{t}0}^{\textup{dr}}$, where
\begin{eqnarray*}
\mu_{\textsc{t}0}^{\textup{dr}}  = E\left[  \varepsilon_0(X) \left\{  e^*(X)  \frac{ 1-Z}{1-e^*(X)}  Y  - \frac{e^*(X) - Z}{1-e^*(X)}  \mu_0^*(X)  \right\}  \right] / e .
\end{eqnarray*}
If $e^*(X) = e(X)$, the conclusion is obvious since the inverse probability weighting estimator is consistent and the additional term has mean zero. If $\mu_0^*(X) = \mu_0(X)$, then by definition, $e\left[\mu_{\textsc{t}0}^{\textup{dr}}  - E\{ Y(0) \mid  Z=1 \} \right]$ equals the expectation of 
$$
\frac{ \varepsilon_0(X)  \mu_0(X)   }{ 1-e^*(X)  }    \Delta_{\textsc{t}}(X)
$$
where
$$
\Delta_{\textsc{t}}(X) = 
 e^*(X) \{1-e(X)\}  - \{  e^*(X) - e(X) \}  - e(X)\{1-e^*(X)\}  = 0.
$$
Therefore, if $\mu_0^*(X) = \mu_0(X)$, then $\mu_{\textsc{t}0}^{\textup{dr}} $ equals $ E\{ Y(1) \mid Z=1\}$. 

\subsection{Proof of Proposition~\ref{prop::bound_constant_epsilons_att}}
Recall that $\varepsilon_{0,\textsc{l}}$ and $\varepsilon_{0,\textsc{u}}$ are the lower and upper bound of $\varepsilon_0(X)$, i.e., $\varepsilon_0(X)\in[\varepsilon_{0,\textsc{l}}, \varepsilon_{0,\textsc{u}}]$. 

First, when $Y\geq 0$, we have $\mu_0(X)=E(Y\mid X,Z=0)\geq 0$. Therefore, $Z\mu_0(X)\geq 0$. By the first identification formula in Theorem~\ref{thm::att-reg-ipw}, we have
\begin{eqnarray*}
    \tau_\T & \geq & E(Y\mid Z=1) - \varepsilon_{0,\textsc{u}}\frac{E\left\{Z\mu_0(X)\right\}}{e}.
\end{eqnarray*}

Second, when $Y\geq 0$, we have $e(X)(1-Z)Y/[e\{1-e(X)\}]\geq 0$. Therefore, by the second identification formula for $\mu_{\T 0}$ in Theorem~\ref{thm::att-reg-ipw}, we have
\begin{eqnarray*}
    \tau_\T &\geq & E(Y\mid Z=1) - \varepsilon_{0,\textsc{u}}E\left[\frac{e(X)(1-Z)Y}{e\left\{1-e(X)\right\}}\right].
\end{eqnarray*}

Again, we prove the third formula by showing the third lower bound is equal to the previous two bounds if either the propensity score or the outcome model is correctly specified. If the propensity score model is correct, we have
\begin{eqnarray*}
    E\left[\frac{\left\{Z-e(X)\right\}\mu_0(X)}{1-e(X)}\right] &=& 0.
\end{eqnarray*}
If the outcome model is correct, we have 
\begin{eqnarray*}
    E\left[e(X)\frac{(1-Z)\left\{Y-\mu_0(X)\right\}}{1-e(X)}\right] &=& 0.
\end{eqnarray*}
Therefore, the third formula of the lower bound is equal to the first two formulas if either the propensity score or outcome model is correct.

The upper bound for $\tau$ can be derived similarly.


\subsection{Proof of Proposition~\ref{prop:semiparametric_bound}}
We compute the semiparametric efficiency bound for $\tau$ and $\tau_\T$ in the following two parts, respectively.
\subsubsection{The semiparametric efficiency bound for $\tau$}

With
\begin{eqnarray}
    \var(\phi_\tau) &=& \var(\phi_1 - \phi_0) \ =\  \var(\phi_1) + \var(\phi_0) - 2\cov(\phi_1, \phi_0), \label{eqn::bound_proof_eqn1}
\end{eqnarray}
we compute the three terms on the right-hand side separately. Let $\mu_z$ denote $E\{Y(z)\}$ for $z=0,1$, we can rewrite the first term in~\eqref{eqn::bound_proof_eqn1} as
\begin{eqnarray*}
    \var(\phi_1) &=& \var(\mathcal{T}_1) + \var(\mathcal{T}_2) - 2\cov(\mathcal{T}_1, \mathcal{T}_2),
\end{eqnarray*}
where $\mathcal{T}_1 = w_1(X)ZY/e(X) - \mu_1$ and $\mathcal{T}_2 = \{Z-e(X)\}\mu_1(X)/\{e(X)\varepsilon_1(X)\}$.
We have
\begin{eqnarray}
    \var(\mathcal{T}_1) &=& E\left\{w_1^2(X) \frac{ZY^2}{e^2(X)} + \mu_1^2 - 2 w_1(X)\frac{ZY}{e(X)}\mu_1\right\} \notag \\
    &=& E\left[\frac{w_1^2(X)}{e(X)}E\left\{Y^2(1)\mid X, Z=1\right\}\right] + \mu_1^2 - 2\mu_1E\left\{w_1(X)\frac{ZY}{e(X)}\right\} \notag \\
    &=& E\left[\frac{w_1^2(X)}{e(X)}\left\{\var(Y\mid X, Z=1) + \mu_1^2(X)\right\}\right] - \mu_1^2, \label{eqn::bound_proof_var_T1} \\
    \var(\mathcal{T}_2) &=& E\left[\frac{\left\{Z-e(X)\right\}^2\mu_1^2(X)}{e^2(X)\varepsilon_1^2(X)}\right] \ =\ E\left\{\mu_1^2(X)\frac{1-e(X)}{e(X)\varepsilon_1^2(X)}\right\},  \label{eqn::bound_proof_var_T2}\\
    \cov(\mathcal{T}_1, \mathcal{T}_2) &=& -E\left[\frac{w_1(X)\mu_1(X)}{e^2(X)\varepsilon_1(X)}Z\left\{1-e(X)\right\}Y\right] \notag \\
    &=& -E\left[\frac{w_1(X)\mu_1^2(X)\left\{1-e(X)\right\}}{e(X)\varepsilon_1(X)}\right] \label{eqn::bound_proof_cov_T1_T2}.
\end{eqnarray}
Combining~\eqref{eqn::bound_proof_var_T1}--\eqref{eqn::bound_proof_cov_T1_T2}, we have
\begin{eqnarray*}
    \var(\phi_1) &=& E\left[\frac{w_1^2(X)}{e(X)}\var(Y\mid X, Z=1) - \mu_1^2 + \left\{e(X)+\frac{1-e(X)}{\varepsilon_1^2(X)}\right\}\mu_1^2(X)\right].
\end{eqnarray*}
Similarly, the second term in~\eqref{eqn::bound_proof_eqn1} equals
\begin{eqnarray*}
    \var(\phi_0) &=& E\left[\frac{w_0^2(X)}{1-e(X)}\var(Y\mid X, Z=0) - \mu_0^2 + \left\{\varepsilon_0^2(X)e(X)+1-e(X)\right\}\mu_0^2(X)\right].
\end{eqnarray*}
The third term in~\eqref{eqn::bound_proof_eqn1} is $\cov(\phi_1, \phi_0)=E\{(\mathcal{T}_1 - \mathcal{T}_2)(\mathcal{T}_3 - \mathcal{T}_4)\}$, where similarly $\mathcal{T}_3 = w_0(X)(1-Z)Y/\{1-e(X)\}-\mu_0$ and $\mathcal{T}_4=\{e(X)-Z\}\mu_0(X)\varepsilon_0(X)/\{1-e(X)\}$. We have
\begin{eqnarray}
    E(\mathcal{T}_1\mathcal{T}_3) &=& 0 - \mu_1E\left\{w_0(X)\frac{(1-Z)Y}{1-e(X)}\right\} - \mu_0E\left\{w_1(X)\frac{ZY}{e(X)}\right\} + \mu_1\mu_0 \ =\ -\mu_1\mu_0, \label{eqn::bound_proof_T1T3} \\
    E(\mathcal{T}_2\mathcal{T}_3) &=& E\left\{w_0(X)\frac{1-Z}{1-e(X)}Y\frac{Z-e(X)}{e(X)\varepsilon_1(X)}\mu_1(X)\right\} \notag \\
    &=& -E\left[\frac{w_0(X)\mu_1(X)}{e(X)\left\{1-e(X)\right\}\varepsilon_1(X)}(1-Z)Y\right] \notag \\
    &=& -E\left\{\frac{w_0(X)\mu_1(X)\mu_0(X)}{\varepsilon_1(X)}\right\}, \label{eqn::bound_proof_T2T3} \\
    E(\mathcal{T}_1\mathcal{T}_4) &=& -E\left[\frac{w_1(X)\mu_0(X)\varepsilon_0(X)}{e(X)}ZY\right] \ =\ -E\left\{w_1(X)\mu_1(X)\mu_0(X)\varepsilon_0(X)\right\}, \label{eqn::bound_proof_T1T4} \\
    E(\mathcal{T}_2\mathcal{T}_4) &=& E\left[\frac{-\left\{Z-e(X)\right\}^2\mu_1(X)\mu_0(X)\varepsilon_0(X)}{e(X)\left\{1-e(X)\right\}\varepsilon_1(X)}\right] \ =\ -E\left\{\frac{\varepsilon_0(X)}{\varepsilon_1(X)}\mu_1(X)\mu_0(X)\right\}. \label{eqn::bound_proof_T2T4}    
\end{eqnarray}
Combining~\eqref{eqn::bound_proof_T1T3}--\eqref{eqn::bound_proof_T2T4}, we have
\begin{eqnarray*}
    \cov(\phi_1, \phi_0) &=& -\mu_1\mu_0 + E\left[\left\{\frac{1-e(X)}{\varepsilon_1(X)}+\varepsilon_0(X)e(X)\right\}\mu_1(X)\mu_0(X)\right].
\end{eqnarray*}
Finally, plugging in the three terms on the right-hand side of~\eqref{eqn::bound_proof_eqn1}, we have
\begin{eqnarray*}
    \var(\phi_\tau) &=& E\left[\frac{w_1^2(X)}{e(X)}\var(Y\mid X, Z=1) - \mu_1^2 + \left\{e(X)+\frac{1-e(X)}{\varepsilon_1^2(X)}\right\}\mu_1^2(X)\right] \\
    &&+ E\left[\frac{w_0^2(X)}{1-e(X)}\var(Y\mid X, Z=0) - \mu_0^2 + \left\{\varepsilon_0^2(X)e(X)+1-e(X)\right\}\mu_0^2(X)\right] \\
    &&+2\mu_1\mu_0 -2E\left[\left\{\frac{1-e(X)}{\varepsilon_1(X)}+\varepsilon_0(X)e(X)\right\}\mu_1(X)\mu_0(X)\right] \\
    &=& E\left\{\frac{w_1^2(X)}{e(X)}\var(Y\mid X,Z=1)+\frac{w_0^2(X)}{1-e(X)}\var(Y\mid X,Z=0)\right\}- \left(\mu_1-\mu_0\right)^2 \\
    && + E\left[e(X)\left\{\mu_1(X)-\mu_0(X)\varepsilon_0(X)\right\}^2 + \left\{1-e(X)\right\}\left\{\frac{\mu_1(X)}{\varepsilon_1(X)}-\mu_0(X)\right\}^2 \right].
\end{eqnarray*}

\subsubsection{The semiparametric efficiency bound for $\tau_\T$} 
In the second part, we derive the semiparametric efficiency bound for $\tau_\T$. The efficient influence function for $\tau_\T$ is
\begin{eqnarray*}
    \phi_{\tau_\T} &=& \frac{Z}{e}\left\{Y-\mu_{\T 1}\right\} - \phi_{\T 0} \ =\ \frac{Z}{e}\left\{Y-\varepsilon_0(X)\mu_0(X)-\tau_\T\right\} + \frac{\varepsilon_0(X)e(X)}{e}\frac{(1-Z)\left\{Y-\mu_0(X)\right\}}{1-e(X)},
\end{eqnarray*}
therefore, we can rewrite the semiparametric efficiency bound for $\tau_\T$ as
\begin{eqnarray*}
    \var(\phi_{\tau_\T}) &=& \var(\mathcal{T}_5) + \var(\mathcal{T}_6) + 2\cov(\mathcal{T}_5, \mathcal{T}_6),
\end{eqnarray*}
where $\mathcal{T}_5 = Z\{Y-\varepsilon_0(X)\mu_0(X)-\tau_\T\}/e$ and $\mathcal{T}_6 = \varepsilon_0(X)e(X)(1-Z)\{Y-\mu_0(X)\}/[e\{1-e(X)\}]$. We have
\begin{eqnarray}
    \var(\mathcal{T}_5) &=& E\left[\frac{Z}{e^2}\left\{Y-\varepsilon_0(X)\mu_0(X)-\tau_\T\right\}^2\right] \notag \\
    &=& E\left(E\left[\frac{Z}{e^2}\left\{Y-\varepsilon_0(X)\mu_0(X)-\tau_\T\right\}^2\mid X\right]\right) \notag \\
    &=& E\left(\frac{e(X)}{e^2}E\left[\left\{Y-\varepsilon_0(X)\mu_0(X)-\tau_\T\right\}^2\mid X,Z=1\right]\right) \notag \\
    &=& E\left(\frac{e(X)}{e^2}E\left[\left\{Y-\mu_1(X) + \mu_1(X) - \varepsilon_0(X)\mu_0(X)-\tau_\T\right\}^2\mid X,Z=1\right]\right) \notag \\
    &=& E\left(\frac{e(X)}{e^2}\left[\var(Y\mid X,Z=1)+\left\{\mu_1(X) - \varepsilon_0(X)\mu_0(X)-\tau_\T\right\}^2 + 0\right]\right) \notag \\
    &=& E\left[\frac{e(X)\var(Y\mid X,Z=1)}{e^2} + \frac{e(X)\left\{\mu_1(X)-\varepsilon_0(X)\mu_0(X)-\tau_\T\right\}^2}{e^2}\right], \label{eqn::bound_proof_var_T5}\\
    \var(\mathcal{T}_6) &=& E\left[\frac{\varepsilon_0^2(X)e^2(X)}{e^2}\frac{(1-Z)\left\{Y-\mu_0(X)\right\}^2}{\left\{1-e(X)\right\}^2}\right] \notag \\
    &=& E\left(E\left[\frac{\varepsilon_0^2(X)e^2(X)}{e^2}\frac{(1-Z)\left\{Y-\mu_0(X)\right\}^2}{\left\{1-e(X)\right\}^2}\mid X\right]\right) \notag \\
    &=& E\left(\frac{\varepsilon_0^2(X)e^2(X)}{e^2\left\{1-e(X)\right\}}E\left[\left\{Y-\mu_0(X)\right\}^2\mid X, Z=0\right]\right) \notag \\
    &=& E\left[\frac{\varepsilon_0^2(X)e^2(X)\var(Y\mid X,Z=0)}{e^2\left\{1-e(X)\right\}}\right], \label{eqn::bound_proof_var_T6} \\
    \cov(\mathcal{T}_5, \mathcal{T}_6) &=& E(\mathcal{T}_5\mathcal{T}_6) = 0. \label{eqn::bound_proof_cov_T5T6}
\end{eqnarray}
Combining~\eqref{eqn::bound_proof_var_T5}--\eqref{eqn::bound_proof_cov_T5T6}, we have the result in Proposition~\ref{prop:semiparametric_bound}.

\end{document}